\documentclass[11pt,a4paper]{article}
\usepackage{times,cite}
\usepackage{graphicx}
\usepackage{amssymb}
\usepackage{amsmath}
\usepackage{color}
\begin{document}

\title{Modelling of magnetic field structure of the Ergodic Divertor of Tore Supra and comparison to the Dynamic Ergodic Divertor of TEXTOR}  

\author{S.S. Abdullaev$^1$, T. Van Rompuy$^2$, K.H. Finken$^1$, M. Lehnen$^1$,
  M. Jakubowski$^1$ \\ \\ $^1$Institut f\"ur Energieforschung$-$4, Forschungszentrum J\"ulich GmbH, \\ EURATOM Association, D-52425 J\"ulich Germany, \\ \\ $^2$Department of Applied Physics, Ghent University \\ Rozier 44, B-9000 Gent, Belgium }

\date{May 9, 2008}

\maketitle

\begin{abstract}
An analytical model previously developed to study the structure of the
magnetic field for the TEXTOR-DED [S.S. Abdullaev et al. Phys. Plasmas, {\bf
  6}, 153 (1999)] is applied to the similar  study of the Ergodic Divertor of
Tore Supra tokamak [Ph. Ghendrih, Plasma Phys. Control. Fusion,  
{\bf 38}, 1653 (1996)]. The coil configuration of ED Tore Supra consists of
six modules equidistantly located along the toroidal direction on the
low-field-side of the torus with given toroidal and poloidal extensions. The
Hamiltonian formulation of field line equations in straight-field-line
coordinates (Boozer coordinates) and the computationally efficient mapping
method for integration of the Hamiltonian field line equations are used to
study the magnetic field structure in the ED. Asymptotical formulas  
for the perturbation magnetic field created by the ED coils are obtained and
the spectrum of magnetic perturbations is analyzed and compared with the one
of the TEXTOR-DED. The structure of ergodic and laminar zones are studied by
plotting Poincar\'e sections, so-called laminar plots (contour plots of wall
to wall connection lengths) and magnetic footprints. The radial profiles of
field line diffusion coefficients are calculated for different perturbation
currents and it is found that for the Tore Supra case in the ergodic zone the
numerical field line diffusion coefficients perfectly follow the quasilinear
formula for smaller perturbation currents although the situation is different
for the maximum perturbation current.
\end{abstract}

\section{Introduction}
The problem of plasma--wall interaction is one of the key issues in nuclear
fusion research. To control the plasma edge, the concept of an 
ergodic divertor has been proposed in
Refs. \cite{EngelhardtFeneberg_78,FenebergWolf_81,Samain_etal_82}). The
idea of the ergodic divertor is based on the creation of a stochastic zone
of magnetic field lines open to the plasma wall which would guide the ionized
particles and energy to special divertor plates, and prevent the
penetration of wall-released impurities into the plasma core. The stochastic
field lines at the plasma edge should be formed by externally created 
resonant magnetic perturbations using external coils.

The ergodic divertor (ED) (ergodic limiter) has been later implemented in
several large size and small size tokamaks. Particularly, an ergodic limiter
has been installed in TEXT (Refs. \cite{Gentle_81,deGrassie_etal_84,%
Ohyabu_etal_84a,Ohyabu_etal_84b,McCool_etal_89,McCool_etal_90}), in Tore Supra
(Refs. \cite{Deschamps_etal_84,Samain_etal_84,Lipa_etal_88,EquipeToreSupra_98,%
Ghendrih_etal_96,Grosman_99,Ghendrih_etal_01,Ghendrih_etal_02}) and in a
number of small-size tokamaks (in JFT-2M (\cite{Shoji_etal_92,Evans_etal_89}),
CSTN-II (\cite{Takamura_etal_87,Takamura_etal_89}), HYBTOK-II
(\cite{Shen_etal_89}), TBR-1 \cite{Caldas_etal_02}, TCABR
(\cite{Pires_etal_05}), and other small fusion devices
(\cite{Kawamura_82,Hattori_etal_84}).

The ergodic divertor implemented in the TEXTOR tokamak, the Dynamic Ergodic
Divertor (DED) \cite{DED_97,Finken_etal_05b,Finken_etal_06b}, has dynamical
features in addition to the conventional ones. It permits the
operation with a rotating magnetic perturbation field, which allows not only
to broaden the heating footprints to avoid hot spots on the divertor plates 
but also to actively influence the plasma rotation and MHD activities
\cite{Finken_etal_04,Koslowski_etal_06}. At the same time the DED perturbation
field has an influence on runaway electron dynamics
\cite{Wingen_etal_06,Finken_etal_07a}, spontaneous density built-up  
\cite{Finken_etal_07b}, and reduction of turbulence transport
\cite{Xu_etal_06}. Particularly, the DED perturbation field has been used to
control edge localized modes (ELM) in limiter H-mode plasmas
\cite{Finken_etal_07a}, similarly to the recent experiments in the DIII-D
tokamak to suppress ELM by the external resonant magnetic perturbations
\cite{Evans_etal_04,Evans_etal_05a,Evans_etal_06}.

The energy and particle transport in the ergodic zone depends significantly on
the structure of the perturbation magnetic field at the plasma edge. Numerous
recent experiments on the TEXTOR-DED have demonstrated that the transport
along the field lines in the presence of the resonant magnetic perturbations
becomes predominant and prevails over the cross field neoclassical and
anomalous diffusion \cite{Jakubowski_etal_06}. The field lines even determine  
a fine structure of heat deposition patterns
\cite{Jakubowski_etal_07a,Jakubowski_etal_07b,Wingen_etal_07}, which is mainly
due to the reduction of the large scale turbulence level at the plasma edge
during DED operation  \cite{Xu_etal_06,Xu_etal_07}. Therefore the study of the
magnetic field structure is an important first step in understanding the
transport properties of the plasma edge modified by the resonant magnetic
perturbations. 

Traditional methods to study the perturbation field structure and its
statistical properties in ergodic divertors have been mainly based on
numerical codes
\cite{Nguyen_etal_95,Nguyen_etal_97,Kaleck_etal_97a,Eich_etal_98}. One of the
approaches to study the formation of the ergodic zone of field 
lines is based on the direct field line tracing in a toroidal system (see,
e.g., \cite{Kaleck_etal_97a}). More in particular, to study the perturbation
magnetic  field structure in the ED of Tore Supra the MASTOC code has been
developed. In its early version \cite{Nguyen_etal_95}, the field lines
equations are formulated in a Hamiltonian form using the intrinsic
coordinates. The Fourier coefficients of the perturbation Hamiltonian are
found by direct integration along the unperturbed field lines using the Biot
and Savart formula and the field line equations are integrated by the
Runge--Kutta integration scheme. 

A new approach to study this problem with an application to the
TEXTOR-DED has been developed in Refs. \cite{Abdullaev_etal_99,%
Finken_etal_99,Abdullaev_etal_03,Finken_etal:2005}. This approach consists of
two steps. The first step is to formulate the equations of magnetic field
lines in Hamiltonian form using the magnetic flux coordinates. For this
purpose the analytical models for the external coil configurations and the
equilibrium magnetic field of the plasma are used \cite{Abdullaev_etal_99,%
Finken_etal_99,Finken_etal:2005}. In the second step the Hamiltonian equations
are integrated using the computationally effective symplectic mapping methods
developed in Refs. \cite{Abdullaev_99,Abdullaev_02,Abdullaev:2006}).

In this work we shall apply these analytical and mapping methods to study the
perturbation magnetic field structure in the ED of Tore Supra.

The paper consists of three sections. In Sect.~\ref{ED_conf} we consider a
model for the ED perturbation coils, analytically calculate the perturbation
magnetic field and study its poloidal and toroidal spectra, as well as its 
radial dependence. There we also study the
poloidal spectrum of the perturbation magnetic field in magnetic flux
coordinates, its relation with the corresponding
spectrum in geometrical coordinates, and compare with the spectrum of
the TEXTOR-DED perturbation field. The structure of the ergodic zone created
by the perturbation field  and the statistical properties of chaotic field
lines are studied in Sect.~\ref{Structure} using  different methods, ranging
from the qualitative Chirikov criterium to the quantitative methods as
Poincar\'e sections, laminar plots and magnetic footprints. In addition, we
calculate the radial profiles of the field line diffusion coefficients. The
summary of obtained results is presented in the conclusive section.

\section{ED coil configuration, perturbation magnetic field and perturbation
  spectra} 
\label{ED_conf}
Tore Supra is a large size tokamak with superconducting toroidal
coils. Its major radius is $R_0=2.38$ m and the minor radius $a=0.8$ m.
The ED coil configuration consists of six identical modules located on the low
field side of the torus and equidistantly spaced along the toroidal angle
$\varphi$ \cite{Samain_etal_84,Nguyen_etal_95}. Each module has a poloidal
extension $\Delta \theta \approx 2\pi/3$ and a toroidal extention $\Delta
\varphi \approx 2\pi/14$. The modules are located at the minor radius $r_c=85$
cm.  

\begin{figure}[htb]
\centering
\hspace{1mm} (a) \hspace{3cm} (b)  \\
\includegraphics[width=4.2cm]{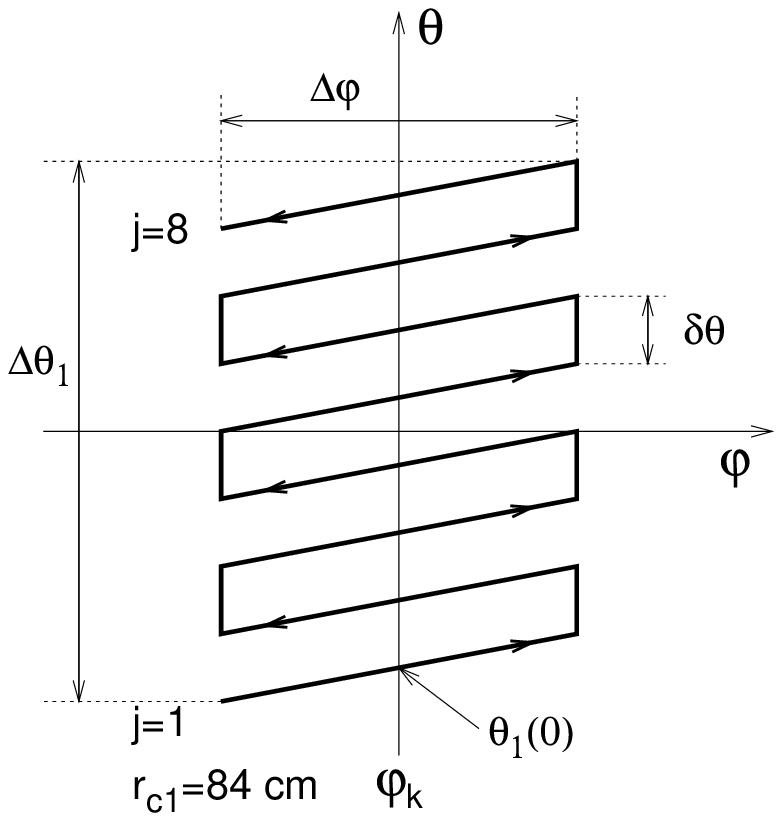}
\includegraphics[width=4.2cm]{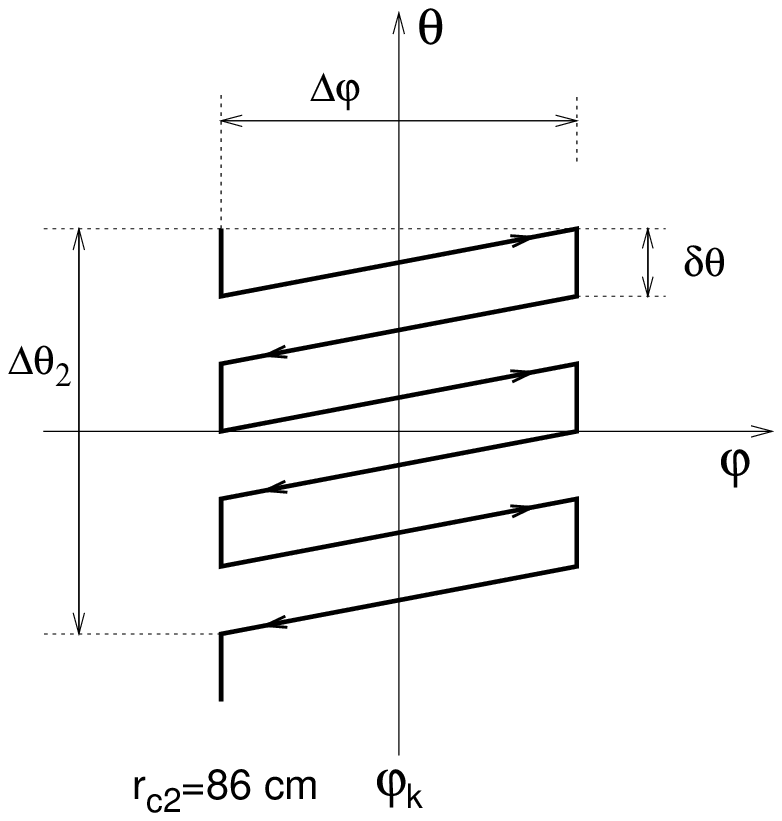}
\caption{Model scheme of one module of the ED coils: (a) the inner winding  at
  $r_{c1} =84$ cm; (b) the outer winding at $r_{c2} =86$ cm.}
\label{Model_ED_coils}
\end{figure}

We model the module with a coil winding shown in Fig. \ref{Model_ED_coils}
where arrows indicate the  current direction. The current flows from the
feeder located at the beginning of the first section $j=1$ of the inner side
of the winding shown in Fig.~\ref{Model_ED_coils}a and returns through the
outer side of the winding shown in Fig.~\ref{Model_ED_coils}b. The minor radii
of the inner and outer sides are $r_{c1} =84$ cm $r_{c2} =86$ cm,
respectively 
\footnote{Furthermore, we use the quasitoroidal coordinate system
$(r,\theta,\varphi)$, which will be referred as geometrical coordinates. }.
Below we calculate the magnetic field, created by the current flowing in this
coil system by considering, in a first time, the case of inner and outer parts
of the winding separately, and by summing those parts afterwards.   

Suppose that the modules are centered near the toroidal angles $\varphi_k =
(k-1) \Delta\varphi$, $\Delta\varphi= 2\pi/6$. The poloidal angle as a
function of the toroidal angle of a point located in the $j-$th
section in the $k-$th module ($k=1, 2, \dots, 6$) is given by   
\begin{eqnarray}  \label{theta_k_phi}
\theta_j^{(k)}(\varphi) = \theta_1^{(k)}(\varphi) + (j-1) \delta \theta,
\hspace{5mm} j = 1, 2, \dots, N ,  
\end{eqnarray}
where $N=8$ for the inner winding and $N=6$ for the outer winding, and $\delta
\theta$ is the poloidal spacing shown in Fig.~\ref{Model_ED_coils}. 
The coordinates of the first coils on each module are described by  
\begin{eqnarray}  \label{theta1_k_phi}
\theta_1^{(k)}(\varphi)=\theta_1(0) + \alpha \left( \varphi-\varphi_k
 \right), \cr \cr  
\mbox{for $\varphi_k - \frac{\Delta \varphi}{2} < \varphi < \varphi_k +
 \frac{\Delta \varphi}{2} $},  
\end{eqnarray}
where $\alpha$ is the slope of a coil creating a helical magnetic
perturbation.  

The poloidal extension, $\Delta \theta$, of a module shown in
Fig.~\ref{Model_ED_coils} can be expressed as a function of the poloidal
position of the first coils, $\theta_1(0)$, at the toroidal section
$\varphi=\varphi_k$: 
\begin{eqnarray}  \label{Delta_theta}
\Delta \theta = 2 |\theta_1(0)| + \alpha \Delta \varphi. 
\end{eqnarray}
 
One should note that this model of coils is not fully equivalent to the Tore
Supra coils. In the latter case the distance between sections of coils in each
module are not equidistant along the poloidal angle $\theta$. It slightly
decreases with the distance from the equatorial plane $\theta=0$
\cite{Ghendrih_95}.

\subsection{Current density}
We describe the current, $I_j$, which flows in a coil section by 
\begin{eqnarray}
I_j^{(i)} = I_d \cos\left(\pi j \right) = (-1)^{j+1}, \hspace{1cm} 
j = 1, 2, \dots, N , 
\end{eqnarray}
where $I_d$ is the current flowing in the coil, $i=1$ for the inner part of
the winding and $i=2$ for its outer part. 

Below we shall consider only the long helical section coils since they
create the magnetic field perturbations that are resonant with the magnetic
field lines of the plasma. The vertical short sections of coils do not
contribute to the resonant field, therefore they will not be taken into
account.

One can introduce the current density vector ${\vec j}(r,\theta,\varphi)$ of
the coil system as 
\begin{eqnarray}  \label{j_mnTS}
{\vec j}_i(r,\theta,\varphi) = {\vec e}^{(i)} \hspace{1mm}
\frac{\delta(r-r_{ci})}{r_{ci}} 
\sum_{k=1}^6 g_\varphi^{(k)} (\varphi) \sum_{j=1}^N I_j^{(i)} \delta
\left(\theta- \theta_j^{(k)}(\varphi) \right),   
\end{eqnarray}
where ${\vec e}^{(i)}= ({\vec e}_r, {\vec e}_\theta,{\vec e}_\varphi)= (0,
\sin\alpha_{0i}, \cos\alpha_{0i})$ is a unit vector 
along the helical section of the coils,  $\alpha_{0i} = \alpha r_{ci}/R_c$,
$R_c=R_0+r_{ci}$,  $(i=1,2)$. Here $g_\varphi^{(k)}(\varphi)$ is a step
function of the toroidal angle $\varphi$ which takes a non-zero value in the
areas covered by coils, i.e.,  
\begin{eqnarray}
g_\varphi^{(k)}(\varphi)= \begin{cases}  1, & \mbox{for $\varphi_k - 
  \frac{\Delta \varphi}{2} < \varphi < \varphi_k + \frac{\Delta
    \varphi}{2} $}, \cr \cr 
0, & \mbox{elsewhere} ,
\end{cases}
\end{eqnarray}
Introducing the step function $g_i(\theta)$ solely depending on the poloidal
angle 
\begin{eqnarray}
g_i(\theta)= \begin{cases} 
1, & \mbox{for $-\frac{\Delta \theta_i}{2} <
    \theta <  \frac{\Delta \theta_i}{2} $}, \cr \cr 
0, & \mbox{elsewhere}, 
\end{cases}
\end{eqnarray}
the current density (\ref{j_mnTS}) can after some transformations be reduced
to  
\begin{eqnarray}   \label{j_sum_3}
{\vec j}_i(r,\theta,\varphi) = {\vec e}^{(i)} \delta(r-r_{ci}) J_0^{(i)} 
g_i(\theta) \sum_{k=1}^6 g_\varphi^{(k)}(\varphi) \cr \times 
\sum_{s=-\infty}^{\infty} \cos \left\{m_0 (2s-1) \left[(\theta-\theta_0)
    -\alpha  (\varphi-\varphi_k)\right ] \right\}.  
\end{eqnarray}
where
\begin{eqnarray} \nonumber
m_0= \frac{\pi}{\delta\theta}, 
\hspace{1cm} J_0^{(i)} = \frac{m_0I_d}{\pi r_{ci}} , \hspace{1cm} 
\theta_0 = \theta_1(0) - \delta\theta. 
\end{eqnarray}

One can show that the current density (\ref{j_sum_3}) can be expanded into a
Fourier series,  
\begin{eqnarray}   \label{j_Fsum_1}
{\vec j}_i(r,\theta,\varphi) = 2 {\vec e}^{(i)} \sum_{m=-\infty}^{\infty}
\sum_{n=-\infty}^{\infty} \sum_{s=1}^{\infty} j_{mn}^{(si)} (r)
\cos\left(m\theta - n\varphi +   \chi_{mn}^{(s)} \right),  
\end{eqnarray}
with the Fourier coefficients 
\begin{eqnarray}   \label{j_F_mn}
j_{mn}^{(si)}(r) = (-1)^q \delta\left(r-r_{ci} \right)
\tilde J_0^{(i)} C_n^{(s)} g_{mi}^{(s)} , \cr \cr 
\chi_{s} = m_0 (2s-1)\theta_0, 
\end{eqnarray}
where 
\begin{eqnarray}  \label{J_0}
{\tilde J}_0^{(i)}=\frac{6J_0^{(i)}\Delta \varphi \Delta \theta_i}{(2\pi)^2} = 
\frac{m_0 I_d}{\pi r_{ci}}\frac{6 \Delta \varphi \Delta \theta_i}{(2\pi)^2},
\\ \label{g_mn}
g_{mi}^{(s)} = \frac{\sin\left([m- m_0 (2s-1)]\Delta\theta_i/2
 \right)}{[m-m_0(2s-1)]\Delta\theta_i/2},
\\ \label{C_n}
C_n^{(s)} = \frac{\sin\left([n- m_0 (2s-1)\alpha] \Delta\varphi/2
 \right)}{[n-m_0(2s-1)\alpha] \Delta\varphi/2}.
\end{eqnarray}

The toroidal mode number $n$ takes values $n=6q, q = 0, \pm 1, \pm 2, \dots$.
As one can see from Eqs. (\ref{j_F_mn}) and (\ref{C_n}), the biggest effect
occurs when the ED coils are designed in such a way that the product $m_0
\alpha$ is close to the toroidal mode $n_0=6$, i.e.,  $|m_0 \alpha - n_0| \ll
\Delta \varphi/2$. Then in the sum (\ref{j_Fsum_1}), the main contribution
comes from the terms with the toroidal numbers $n=(2s-1)n_0$, ($s=1,2,
\dots$). Leaving in Eq. (\ref{j_Fsum_1}) only these terms we have 
\begin{align}   \label{j_Fsum_2s_res}
{\vec j}_i(r,\theta,\varphi) &= 2 {\vec e}^{(i)} \sum_{m=-\infty}^{\infty}
\sum_{s=1}^{\infty} j_{mn}^{(si)}(r) \times \cos \left(m\theta -
  (2s-1)n_0 \varphi +   \chi_{s} \right),  
\end{align}
where $n_0 = 6$, $i=1,2$. 

\subsection{Magnetic field in a cylindrical approximation}
First we consider the magnetic field created by the helical currents
(\ref{j_Fsum_2s_res}). Using the procedure similar in
Ref. \cite{Finken_etal:2005} one can obtain the exact formula for
the scalar potential $\Phi(r,\theta,\varphi)$. Then the perturbation magnetic
field is given by ${\bf B}(r,\theta,\varphi) = \nabla\Phi(r,\theta,\varphi)$. 
For $r < r_{ci}, (i=1,2)$ one has
\begin{align}   \label{F_mns}
\Phi(r,\theta,\varphi) &= \Phi_1(r,\theta,\varphi) + \Phi_2(r,\theta,\varphi),
\cr 
\Phi_i(r,\theta,\varphi) &= \sum_{m=-\infty}^{\infty} \sum_{n}
\Phi_{mn}^{(i)}(r) 
\sin \left(m\theta - n \varphi+  \chi_{s} \right),  
\end{align}
where $n=(2s-1)n_0$, ($s=1,2, \dots$) and 
\begin{align}    \label{Fi_mns}
 \Phi_{mn}^{(i)}(r) &= - B_c^{(i)} C_n^{(s)} g_m^{(si)} f_{mn}^{(i)}(r)
  \frac{r_{ci}}{m},   \cr      
f_{mn}^{(i)}(r) &= -\frac{2nr_{ci}}{R_{ci}} K'_m\left(\frac{nr_{ci}}{R_{ci}}
  \right)  I_m\left(\frac{nr}{R_{ci}} \right) ,    \cr   
B_c^{(i)} &= \frac{2 \mu_o m_0 I_d \cos (\alpha_{0i})}{\pi r_{ci}}
 \frac{6 \Delta\varphi \Delta\theta_i}{(2\pi)^2},  
\end{align}
where $I_m(z)$ and $K_m(z)$ are the modified Bessel functions ($K_m'(z) \equiv
dK_m(z)/dz$).  

Here $B_{c}^{(i)}$ is the characteristic amplitude of external magnetic field
strength \footnote{We should note that the definition of $B_{c}^{(i)}$ as well
as coefficients $g_m^{(si)}$ are slightly different from the corresponding ones
 given in Ref. \cite{Abdullaev_etal_99,Finken_etal:2005}}.  
For the typical parameters of the ED of Tore Supra ($r_c = 0.85$ m, 
$I_d=$ 22.5 kA) we have $B_c \approx 425$ G. It is close to the value, $B_c
\approx 405$ G, of the corresponding quantity for the TEXTOR-DED (at $I_d=15$
kA).   

\begin{figure}[t]
\centering
\hspace{1cm} (a) \hspace{4cm} (b) \\
\includegraphics[width=5cm]{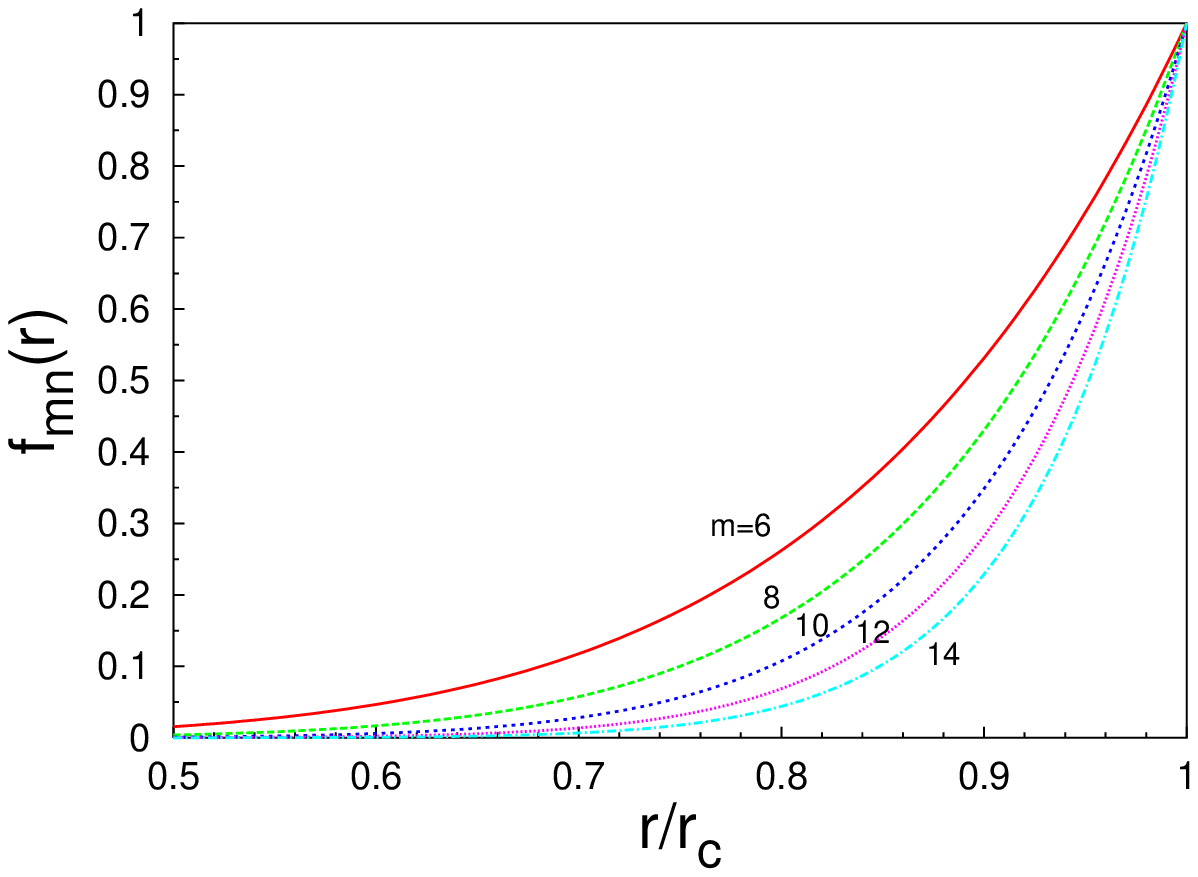}
\includegraphics[width=5cm]{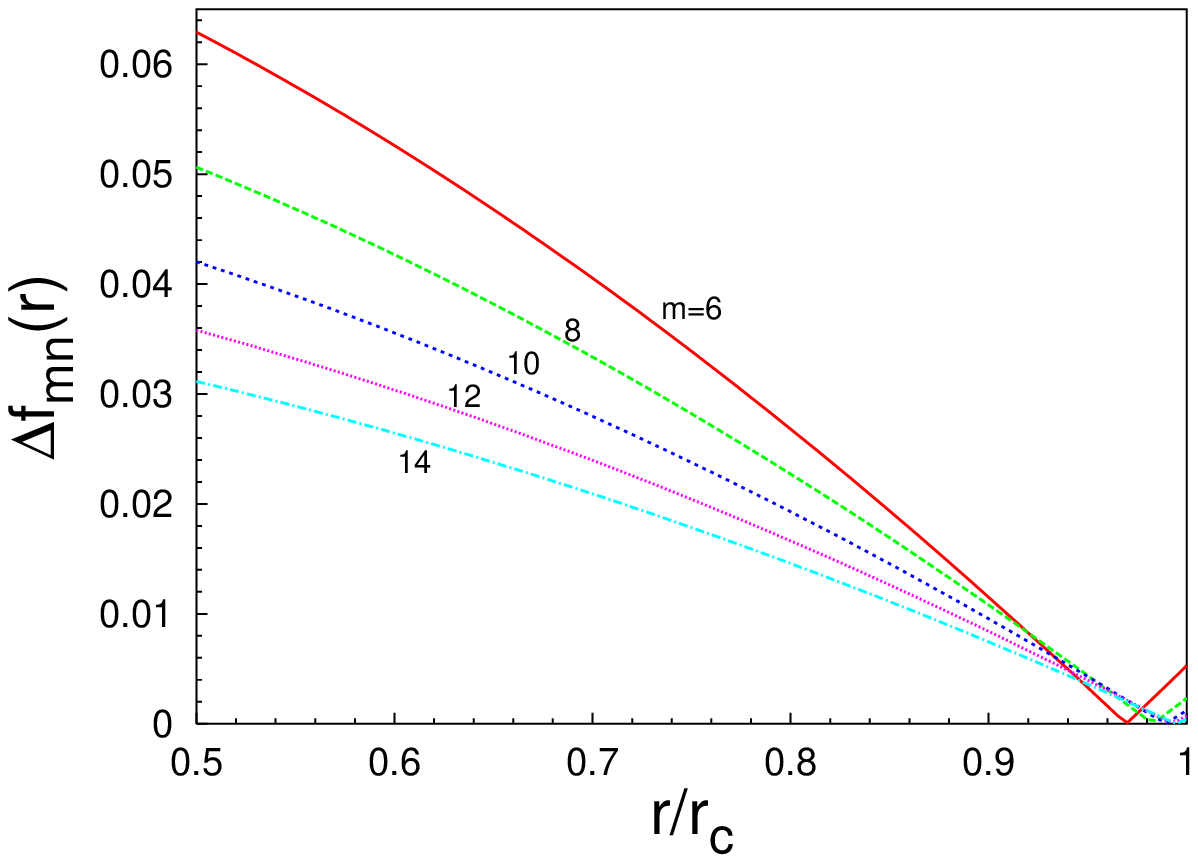}
\caption{(a) Radial dependence of the function $f_{mn}(r)$ for different
  poloidal modes $m$. (b) Relative deviation of the $f_{mn}(r)$ from the power
  law $(r/r_c)^m$: $\Delta f_{mn}(r) = |f_{mn}(r)-(r/r_c)^m|/f_{mn}(r)$. 
  Parameters are $r_c=85$ cm, $R_0=238$ cm, $n=6$. } 
\label{f_mn}
\end{figure}

The radial dependences of the perturbation field are described by functions
$f_{mn}^{(i)}(r)$ which are shown in Fig.~\ref{f_mn} for a several mode numbers
$m$. For large mode number $m$  ($m \geq 4$) the radial dependence is well
described by the following asymptotical formula \(f_{mn}^{(i)}(r) \approx
\left( r/r_{ci} \right)^m \). Then the radial component of the magnetic field
$B_r$ can be represented as 
\begin{align}   \label{Br_mns}
B_r(r,\theta,\varphi) &= B_r^{(1)}(r,\theta,\varphi) + B_r^{(2)}
(r,\theta,\varphi), \cr 
 B_r^{(i)}(r,\theta,\varphi) &= - \sum_{m=-\infty}^{\infty} \sum_{s=1}^{\infty}
B_c^{(i)} C_n^{(s)} g_m^{(si)} \left( \frac{r}{r_{ci}} \right)^{m-1} 
\sin \left(m\theta - (2s-1) n_0 \varphi+  \chi_{s} \right).  
\end{align}

\subsection{Toroidal corrections}
We consider the first order correction to the magnetic field perturbations
due to toroidicity of the system. According to \cite{MorozovSolovev_66} this
effect can be taken into account by multiplying the scalar potential
$\Phi(r,\theta,\varphi)$ obtained in the cylindrical approximation with the
factor $\sqrt{R_0/\left(R_0+r\cos\theta \right)}$ when the  corrections
of order $\left(nr_c/2R_c \right)^m$ are small. In this case we have
\begin{eqnarray}   \label{F_mns_tor}
\Phi(r, \theta, \varphi) = \sqrt{\frac{R_0}{R_0+r\cos\theta}}
\left[\Phi_1(r, \theta, \varphi) + \Phi_2(r, \theta, \varphi) \right], 
\end{eqnarray}
where the functions $\Phi_i(r, \theta, \varphi)$, $(i=1,2)$ are given by
Eq. (\ref{F_mns}). We present the corresponding radial magnetic field as a sum
of contributions corresponding to different toroidal modes, $n$, 
\begin{align}   \label{B_r_tor}
B_r(r, \theta, \varphi)&=  \frac{\partial \Phi}{\partial r} = \sum_n
B_r^{(n)}(r, \theta, \varphi), \\ 
B_r^{(n)}(r, \theta, \varphi) &= \sum_{m=-\infty}^{\infty} B_{mn}(r,\theta)
\sin \left( m\theta - n\varphi + \chi_s\right) , \nonumber
\end{align}
where 
\begin{align}   \label{Bn_r_tor}
B_{mn}(r,\theta) &= \frac{\partial }{\partial r}
\sqrt{\frac{R_0}{R_0+r\cos\theta}} \left[\Phi_{mn}^{(1)}(r) +
  \Phi_{mn}^{(2)}(r) \right] \cr 
& = \left[ B_c^{(1)} g_{m1}^{(s)} \left( \frac{r}{r_{c1}}
\right)^{m-1} + B_c^{(2)} g_{m2}^{(s)} \left( \frac{r}{r_{c2}} \right)^{m-1}
\right] \cr & \times \sqrt{\frac{R_0}{R_0+r\cos\theta}} \left( 
1- \frac{r \cos\theta}{ 2m \left(R_0+r\cos\theta \right)} \right). 
\end{align}

The toroidal component of the vector potential of the perturbation
field takes the form 
\begin{align}   \label{A_mns_tor}
&A_\varphi(r, \theta, \varphi) = 
\epsilon B_0 R_0 a(r, \theta, \varphi) =
\epsilon B_0 R_0 \sum_{m=-\infty}^{\infty} \sum_{n} a_{mn}(r,\theta) \cos
\left(m\theta - n   \varphi+  \chi_{mn} \right), \cr
& a_{mn}(r,\theta) = m^{-1} r B_{mn}(r,\theta),   
\end{align}
where $\epsilon$ stands for the dimensionless perturbation parameter, defined
as $\epsilon = B_c^{(1)}/B_0$, $B_0$ is the strength of the toroidal
field. The dimensionless Fourier coefficients, $a_{mn}(r,\theta)$, are then
given by
\begin{align} 
a_{mn}(r,\theta)=  a_{mn}^{(1)}(r,\theta) + \delta a_{mn}^{(2)}(r,\theta), 
\end{align}
where 
\begin{align}   \label{a_mn_i}
a_{mn}^{(i)}(r,\theta) &= g_m^{(i)} \frac{r_{ci}}{mR_0} \frac{r}{m}
\frac{d}{dr} \left( \sqrt{\frac{R_0}{R_0+r\cos\theta}} \left( \frac{r}{r_{ci}}
  \right)^{m} \right)   \cr
&= g_{mi}^{(s)} \frac{r_{ci}}{mR_0} \left( \frac{r}{r_{ci}} \right)^{m}
\sqrt{\frac{R_0}{R_0+r\cos\theta}} \left(
1- \frac{r \cos\theta}{ 2m \left(R_0+r\cos\theta \right)} \right). 
\end{align}
Here $(i=1,2)$, $\delta= B_{c}^{(2)}/B_{c}^{(1)} = r_{c1}\Delta\theta_2/
\left(r_{c2} \Delta\theta_1 \right)$. The phase $\chi_{mn} = \chi_s$ and
toroidal mode number $n = (2s-1)n_0$. 

\begin{figure}[t]
\hspace{5cm} (a) \hspace{6cm} (b) \\
\includegraphics[width=6cm]{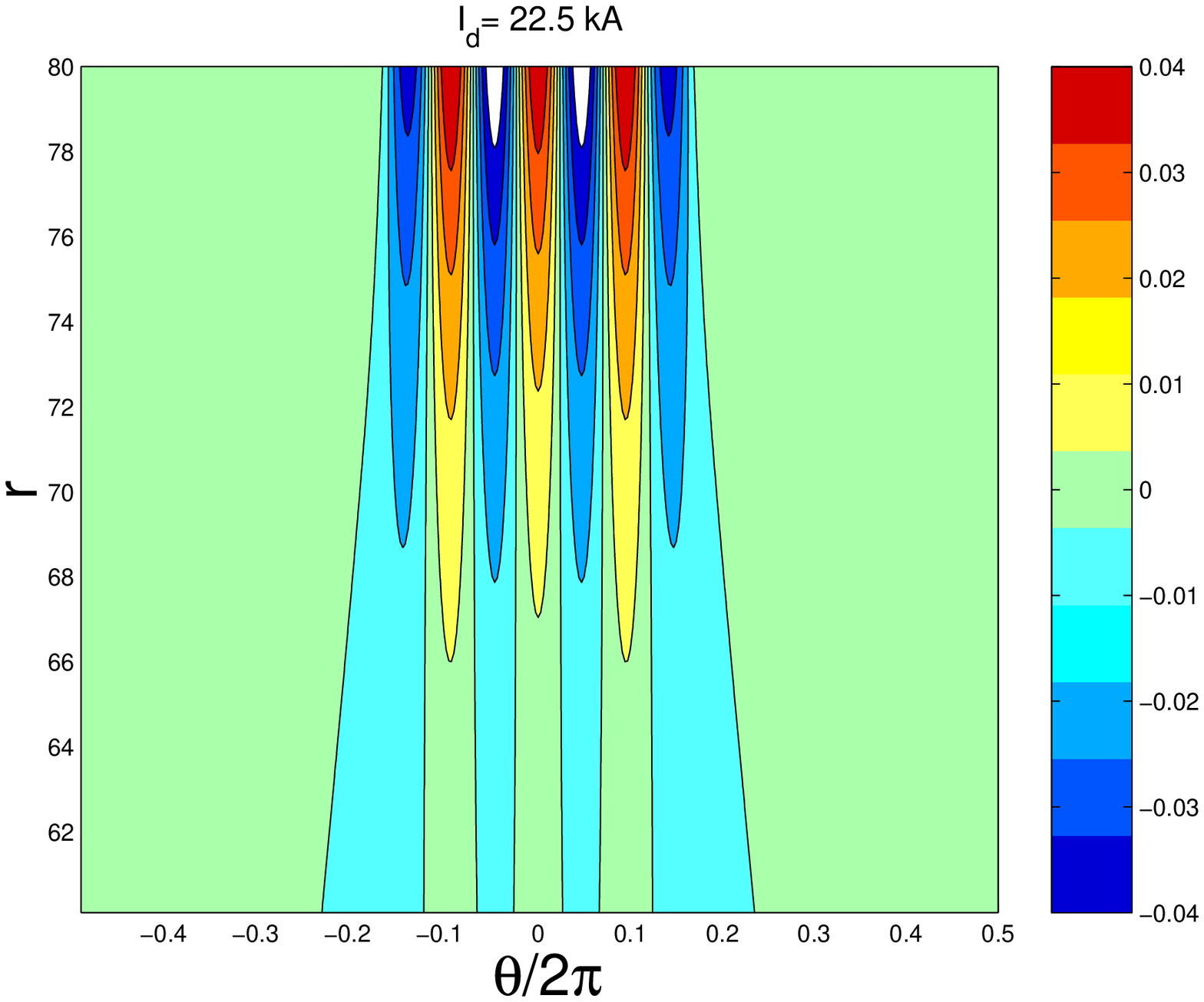} 
\includegraphics[width=7cm]{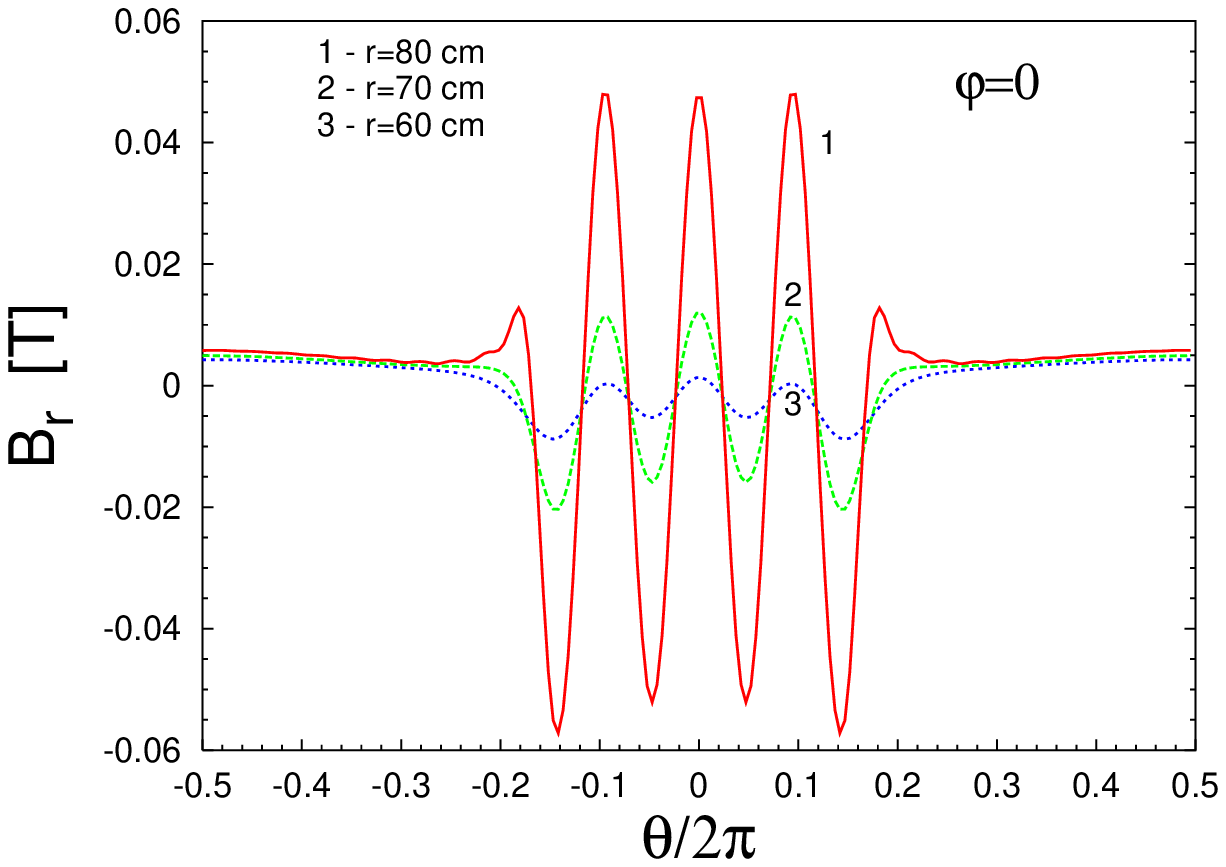}
\caption{Radial component of the perturbation field $B_r$: (a) contour plot 
  in the ($\theta,r$) plane ($\varphi=0$); (b) poloidal dependence at different radial positions 
$r$: 1 $-$ $r$= 80 cm , 2 $-$ $r$= 70 cm, 3 $-$ $r$= 60 cm. $I_d$= 22.5 kA. }
\label{Br_mn}
\end{figure}

The angular dependencies of the perturbation field $B_r(r,\theta,\varphi)$ 
at the fixed values of radial coordinate $r$ and the toroidal angle $\varphi=0$
are plotted in Fig. \ref{Br_mn}.
The radial dependence of the perturbation field is determined by the poloidal
mode spectra, $g_{mi}$. According to the definition given by Eq. (\ref{g_mn})
the latter is localized near the central mode $m_c = (2s-1)m_0 = nm_0/n_0$,
($s=1,2, \dots$) and has a width $\Delta m = \pi/\Delta \theta_i$. Therefore,
one expects that the perturbation field, $B_r^{(n)}(r, \theta, \varphi)$ of
the $n-$th toroidal mode has the following radial dependence,
\begin{eqnarray}  \label{B_r_gamma}
B_r^{(n)}(r,\theta,\varphi) \propto \left( \frac{r}{r_{ci}}
\right)^{\gamma_{n}}, \hspace{6mm}  \gamma_{n}= \frac{nm_0}{n_0}-1. 
\end{eqnarray}
The power law of the radial decay of perturbation field has the lowest
exponent, $\gamma_{n=6} =m_0-1$, for the toroidal mode $n=6$ . For the value
$\delta \theta =18^{\circ}$, one has $m_0= \pi/ \delta \theta = 10$ the
exponent $\gamma = 9$. For the next toroidal mode $n=18$ we have
$\gamma_{n=18} = 3m_0-1 = 29$.

It should be pointed out that the power--law decay of the perturbation field 
(\ref{B_r_gamma}) found here is different from the exponential decay law 
$\exp(-m_0(r_{ci}-r)/r_{ci})$ proposed for the ED in Tore Supra
\cite{Nguyen_etal_95}. 

\subsection{Determination of the spectra of magnetic perturbation
  $H_{mn}(\psi)$ }\label{Determ_Spectra}

Starting from the expressions (\ref{B_r_tor}) to (\ref{a_mn_i}) for the
perturbation field $B_r(r,\vartheta,\varphi)$ and the toroidal component of
its vector potential, the field line structure in the Tore Supra ED edge can
be studied following the approach described in \cite{Finken_etal:2005}. In
terms of magnetic flux coordinates ($\psi_{pol}, \psi \equiv \psi_{tor},
\vartheta, \varphi$) the magnetic field ${\bf B}$ can be presented in the
Clebsch form \cite{Boozer_83,Balescu:1988,Boozer_92}, where ($\psi_{pol},
\psi$) are poloidal and toridal fluxes, ($\vartheta, \varphi$) are poloidal
and toroidal angles. In this coordinate system the field line equations have
the Hamiltonian form 
\begin{eqnarray}   \label{H_eqn}
\frac{d\psi}{d\varphi} = - \frac{\partial \psi_{pol}}{\partial \vartheta}
\hspace{1cm} 
\frac{d\vartheta}{d\varphi} =  \frac{\partial \psi_{pol}}{\partial \psi}, 
\end{eqnarray}
with the  Hamiltonian
\begin{align}   \label{H_field}
& \psi_{pol} = \psi_{pol}^{(0)}(\psi) + \psi_{pol}^{(1)}(\psi, \vartheta,
\varphi) , \cr 
&\psi_{pol}^{(0)}(\psi) = \int \frac{d\psi}{q(\psi)}, \cr
&\psi_{pol}^{(1)}(\psi, \vartheta, \varphi) = \epsilon \sum_{m,n} H_{mn}(\psi)
\cos(m\vartheta - n \varphi+  \chi_{mn}).    
\end{align}
where $q(\psi)$ is the safety factor of the equalibrium plasma, $H_{mn}(\psi)$
are the Fourier coefficients of the perturbation field corresponding to the
poloidal mode numbers $m$ and the toroidal mode $n$. The equations
(\ref{H_eqn})  are integrated using the computationally efficient mapping
method (see Refs. (\cite{Abdullaev_99,Abdullaev_02,Abdullaev:2006}).  

The equilibrium plasma of the Tore Supra tokamak is modelled by a plasma with
nested, circular magnetic surfaces with an outward (Shafranov) shift due to
effects of the plasma pressure and electric current, again completely similar
to the equilibrium  
plasma model described in \cite{Finken_etal:2005} (which is based on
expressions for the Shafranov shift, the toroidal current density profile of a
cylindrical plasma model and the representation of the safety factor due to
toroidicity as a series of powers of the inverse aspect ratio given in
\cite{Nguyen_etal_95}, \cite{Wesson:2004} and \cite{Abdullaev_etal_99},
respectively). The ripple of the magnetic field was not yet incorporated in
the modelling presented here, thus ruling out the possibility of a direct
comparison of the results with experimental evidence. 

In order to integrate the field line equations, the Fourier components
$H_{mn}(\psi)$ of the perturbation Hamiltonian have to be determined. 
According to Refs. \cite{Abdullaev_etal_99,Abdullaev:2006,Finken_etal:2005}
those are found by the following Fourier integral from the product of the
major radius $R=R_0+r\cos\theta$ and the vector potential $A_\varphi$ of the
perturbation field (\ref{A_mns_tor}), 
\begin{eqnarray}   \label{H_mn}
  H_{mn}(\psi) = \mbox{Re } \iint\limits_0^{2\pi} \frac{RA_\varphi
 (r,\theta, \varphi)}{(2\pi)^2B_0R_0^2} e^{-im\vartheta+in\varphi} d\vartheta
  d\varphi,  
\end{eqnarray}
where $(r,\theta)$ is the unperturbed field line on the given magnetic surface
$\psi$ and $\vartheta$ is the intrinsic poloidal angle which is different from
the geometrical poloidal angle $\theta$. In the Eq. (\ref{H_mn}) it is
supposed that $(r,\theta)$ is a function of the $\psi$ and $\vartheta$:
$r=r(\psi,\vartheta), \theta = \theta(\psi,\vartheta)$. 

Using the Fourier series in Eq. (\ref{A_mns_tor}) the integral can be
reduced to 
\begin{align}   \label{H_mn_a_mn}
H_{mn}(\psi) = \frac{1} {2\pi}\sum_{m'} \int\limits_0^{2\pi} b_{m'n} (r,\theta)
  e^{-i[m\vartheta - m'\theta(\vartheta)]} d\vartheta, \cr
  b_{m'n} (r,\theta) = a_{m'n} (r,\theta) \left(1 + \frac{r\cos\theta}{R_0}
  \right)^{1/2}.  
\end{align}
As seen from Eq. (\ref{H_mn_a_mn}) the toroidal spectra, $n$, of the
perturbation field remain unchanged. However, the poloidal spectra,
$b_{mn}(r,\theta)$, in the geometrical space, $(r,\theta)$, changes
drastically in the $(\psi,\vartheta)$ space. First following
Refs. \cite{Abdullaev_etal_99,Abdullaev:2006,Finken_etal:2005} we
qualitatively estimate the integral (\ref{H_mn_a_mn}) in order to reveal the
main features of the transformation of the spectra of magnetic perturbations.

The integral (\ref{H_mn_a_mn}) contains the fast oscillating phase
$f(\vartheta) = m\vartheta - m'\theta(\vartheta)$ and slowly varying amplitude
$b_{m'n}(r,\theta)$. Therefore it can be estimated using asymptotical methods
\cite{Fedoryuk:1989}. As was shown in \cite{Abdullaev:2006,Finken_etal:2005}
for the perturbation field located on the low field side of the torus the
$H_{mn}(\psi)$ is given by the following asymptotical formula  
\begin{align}   \label{H_mn_a_mn_asym}
  H_{mn}(\psi) &= \sum_{m'} b_{m'n} \left(r,\theta=0\right)
  \left(\frac{2}{|\gamma_3|m'} \right)^{1/3}  \mbox{Ai } \left( -
  \frac{\gamma_1 m' - m}{\left(|\gamma_3|m'/2 \right)^{1/3}} \right), 
\end{align}
where the coefficients $\gamma_1$ and $\gamma_3$ are the first and third
derivatives of the geometrical angle $\theta$ with respect
to the intrinsic angle $\vartheta$ taken on the low field side of the torus,
$\theta=0$, respectively:  
\begin{eqnarray}   \label{theta_der}
  \gamma_1= \frac{d\theta}{d\vartheta}\bigg|_{\theta=0}, \hspace{1cm} 
  \gamma_3= \frac{d^3\theta}{d\vartheta^3} \bigg|_{\theta=0}.  
\end{eqnarray}
We should note that $\gamma_1>1$, $\gamma_3<0$, and their particular values
depend on the equilibrium plasma configuration. The sum (\ref{H_mn_a_mn_asym})
can be replaced by integration over $m'$ and asymptotically estimated. This
procedure gives
\begin{align}   \label{H_mn_asym}
  H_{mn}(\psi) \approx \gamma_1^{-1} B b_{m'(m) n}\left(r,\pi=0\right), 
\end{align}
where $m'(m)$ is determined by  
\begin{eqnarray}    \label{mn_asym}
  m'(m)= \frac{m- x_c \left(|\gamma_3|m/2\right)^{1/3}}{\gamma_1},
\end{eqnarray}
and $B= \sqrt{2\pi/|x_c|} \mbox{Ai }(x_c)$, and $x_c \approx -1$ is the local
maximum of the Airy function.

The obtained relation (\ref{H_mn_asym}) describes the transformation law of
the poloidal spectra of magnetic perturbations in a geometrical space to the
ones in intrinsic coordinates. As was mentioned above the poloidal mode spectra
$a_{mn}$ at the given toroidal mode $n$ are localized near the central mode
$m_c = nm_0/n_0$ with a poloidal mode extension $\Delta m = \pi/\Delta
\theta_i$. Then from Eqs. (\ref{H_mn_asym}), (\ref{mn_asym}) it follows that
the spectrum, $H_{mn}$, has the same form as $a_{mn}$, but its central mode
$m_c^*$ is shifted to the higher number $m_c^* \approx m_c \gamma_1$ and the
width becomes larger $\Delta m^* \approx \Delta m\gamma_1$, since $\gamma_1
>1$. These rigorously obtained results coincide with the corresponding ones
obtained in \cite{Ghendrih_95,Nguyen_etal_95} by a qualitative analysis. 

However, the radial dependence of perturbation modes $H_{mn}(\psi)$
(\ref{H_mn_asym}) has a power-law $H_{mn}(\psi) \propto r^{m/\gamma_1}$ in
contrast to the exponential law $\exp\left(-m(r_c-r)/r_c \right)$ supposed in
\cite{Nguyen_etal_95,Nguyen_etal_97} . 

\begin{figure}[t]
  \centering
  \includegraphics[width=8cm]{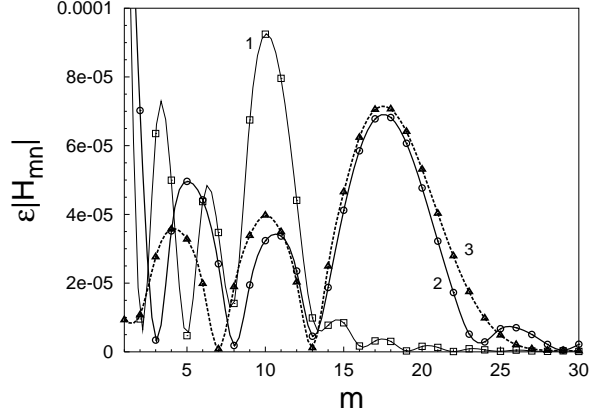}
  \caption{Poloidal spectra of magnetic perturbation field at the magnetic
    surface $q=3$: Curve 1 describes $b_m(r,0)$, curve 2 corresponds to the
    asymptotical formula (\ref{H_mn_asym}), (\ref{mn_asym}) with the fitting
    parameters $B=1.4$, $x_c=-0.5$, and curve 3 corresponds to the numerical
    calculations of the integral (\ref{H_mn_a_mn}). The plasma parameters are
    the plasma minor radius $a=80$ cm, the major radius $R_0=238$ cm, the
    toroidal field $B_t=3.03$ T, the plasma current $I_p$=1.5 MA, the plasma
    $\beta_{pol}=0.13$. The ED current $I_d$= 22.5 kA, $\delta\theta =
    19^{\circ}$, $\Delta\varphi=\pi/14$. The quantities $\gamma_1 = 1.85$,
    $\gamma_3 =-3.124$. }   
  \label{asymp_spectra}
\end{figure}

The mode transformation is illustrated in Fig.~\ref{asymp_spectra} where the
comparison of the asymptotical formula (\ref{H_mn_asym}), (\ref{mn_asym}) for
$H_{mn}(\psi)$ (curve 2) with its numerically calculated value (curve 3) from
the integral (\ref{H_mn_a_mn}) is shown. Parameters $B$ and $x_c$ in the
asymptotical formula are considered as fitting parameters. Curve 1 corresponds
to the spectra $b_m(r,0)$ in the geometrical space. As seen from
Fig.~\ref{asymp_spectra} the asymptotical formula describes well the
transformation law of the poloidal spectra of magnetic perturbations.


The asymptotical formula (\ref{H_mn_asym})
resulting from  the qualitative analysis of the transformation of perturbation
modes in geometrical coordinates to the ones in flux coordinates is important,
as it contributes to understanding the features of mode transformations in
toroidal systems and clearly shows the differences between the cases when the
perturbation coils are located on the low-field-side (for the ED of Tore
Supra) and on the high-field side (the TEXTOR-DED, for which the corresponding
formula is given in Sect.~\ref{Structure_compareTEXTOR}), 
without having to resort to numerical computations.

\begin{figure}[htb]
\hspace{5cm} (a) \hspace{6cm} (b)  \\
  \includegraphics[width=6.8cm]{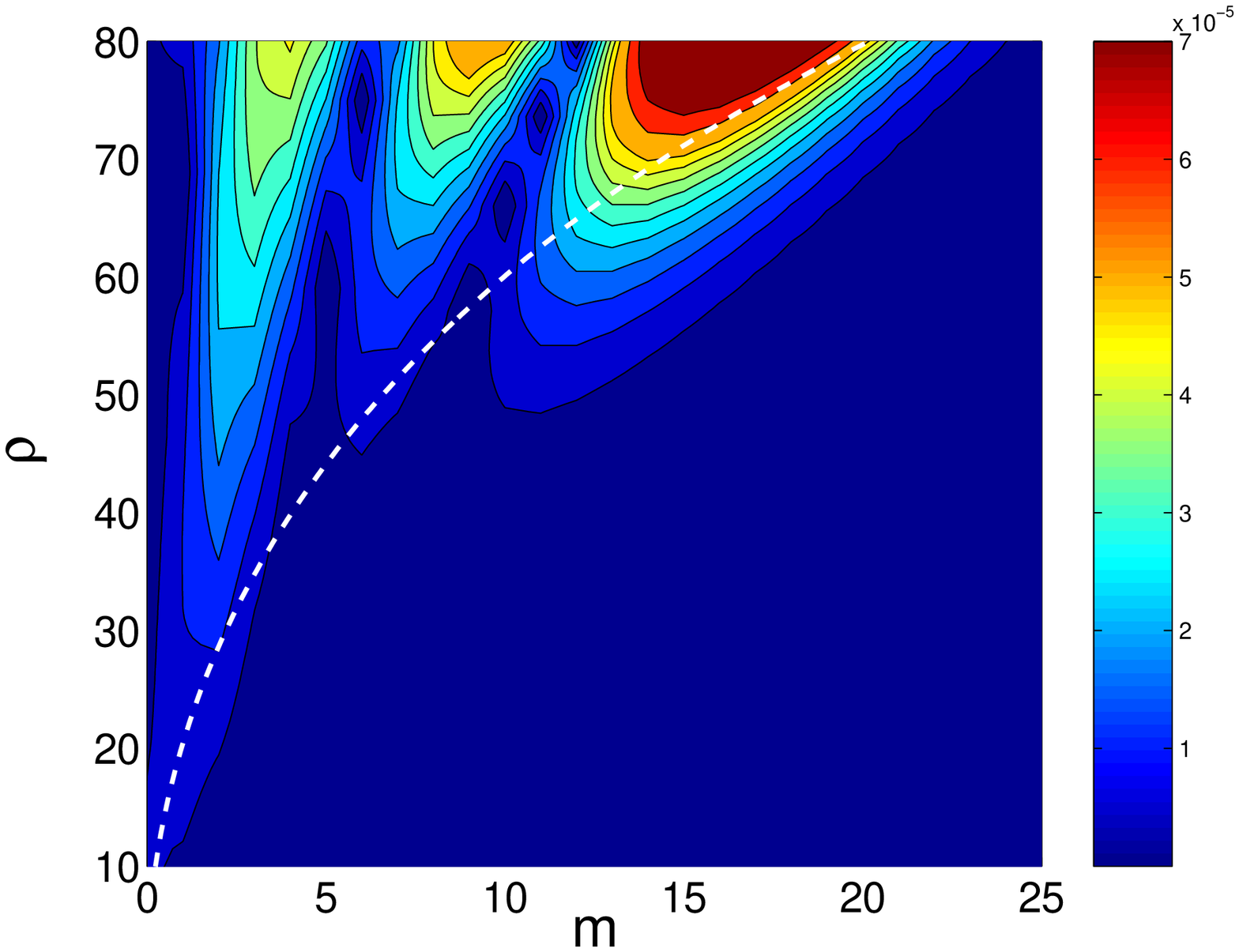} 
  \includegraphics[width=6.6cm]{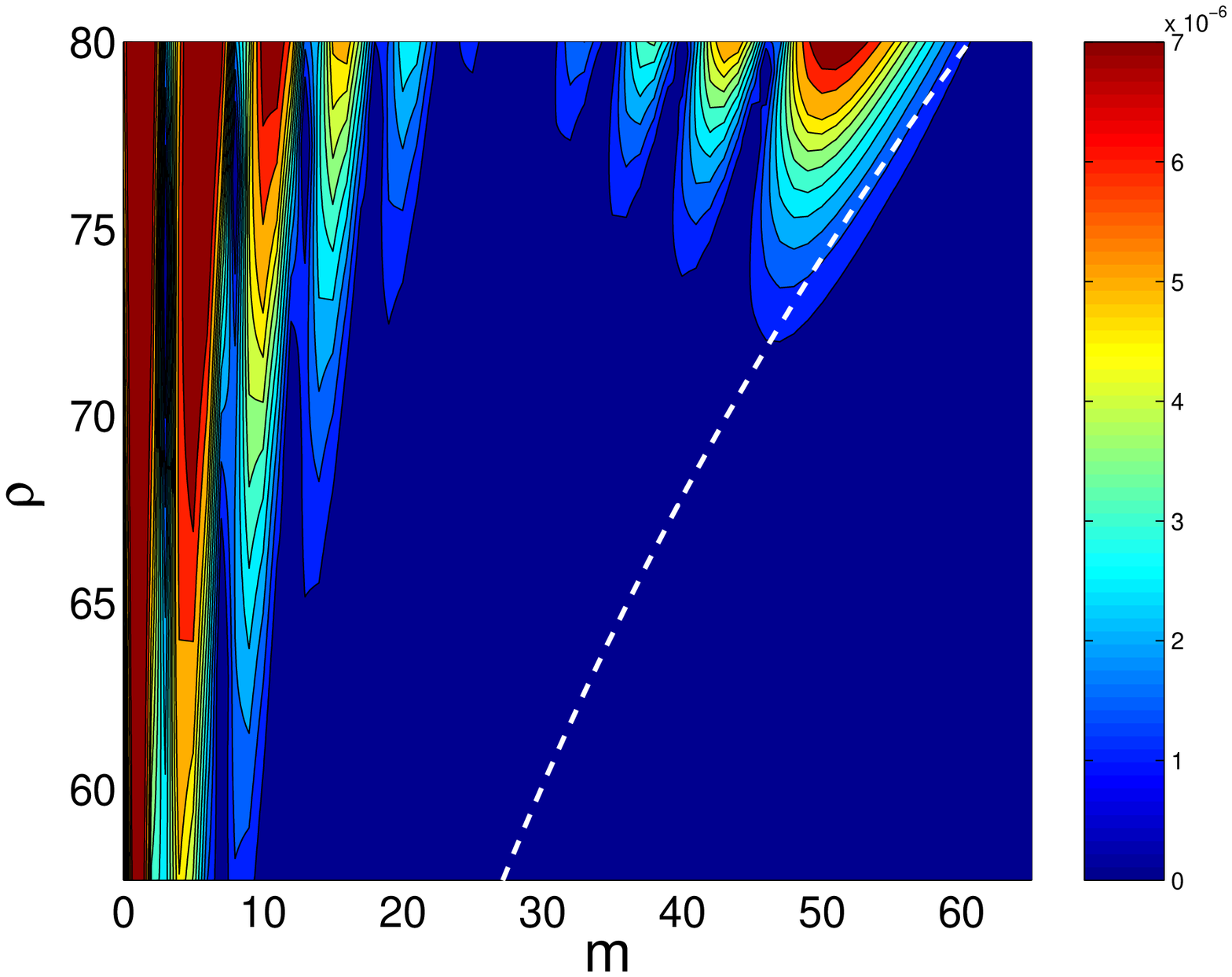}
\caption{(a) Contour plot of poloidal spectrum of perturbation field, $\epsilon
    |H_{mn}|$, in the ($m,\rho)-$ plane for the toroidal mode $n=6$. White
    dashed curve corresponds resonant line $(nq(\rho),\rho)$.  
    (b) The same as in (a) but for the toroidal mode $n=18$. Parameters are
    the same as in Fig. \ref{asymp_spectra}. }    
  \label{H_mnIp1500}
\end{figure}

Fig.~\ref{H_mnIp1500} shows the contour plot of the poloidal spectra of
magnetic perturbations $H_{mn}(\psi)$ for the toroidal modes $n=6$ and $n=18$,
respectively. The values $H_{mn}(\psi)$ at the resonant surfaces $\rho_{mn}$
(or $\psi_{mn}$, $nq(\psi_{mn})=m$) lie at the white curve $nq(\rho)$.    
One can see that the magnetic perturbation have a wide spectrum with the
central modes $m_c^*(\psi_{mn})$ at the rational magnetic surface $\psi_{mn}$,
($q(\psi_{mn})=m/n$) close to the corresponding resonant mode numbers $m$. 
It covers the rational magnetic surfaces located at the plasma edge between $q
\geq 2$ and $q=3.5$. The magnetic perturbation for the toroidal mode $n=18$ is
one order smaller than for the mode $n=6$. 

The resonant components of $H_{mn}(\psi_{mn})$ decay inwardly as
$H_{mn}(\psi_{mn}) \propto \psi_{mn}^{m_r/2}$, with $m_r \approx 7.9$ for the
toroidal mode $n=6$.

\subsection{Comparison with the TEXTOR-DED}\label{Structure_compareTEXTOR} 

Unlike the ED coils in Tore Supra, the perturbation coils of the DED of
TEXTOR are located on the high-field-side (HFS) of the torus
\cite{FinkenWolf_97,Abdullaev_etal_99,Finken_etal_99,Finken_etal:2005}. It
consists of 16 helical coils continuously winding around the torus with the
poloidal extension, $\Delta \theta \approx 72^{\circ}$. By special arrangement
of the current distribution over the coils one can create the perturbation
field with the dominant toroidal modes, $n=4$, $n=2$, and $n=1$, which are
called $m:n$=12:4, 6:2, and 3:1 operational modes, respectively. The poloidal
spectrum, $|b_{mn}(r,\theta)|$, of the perturbation field in the geometrical
space has a wide spectrum with the central mode $m_c=n m_0/4 \approx 5n $ and
the width $\Delta m \approx \pi/\Delta \theta$, where $m_0\approx 20$. 
\begin{figure}[htb]
  \centering 
  \includegraphics[width=8cm]{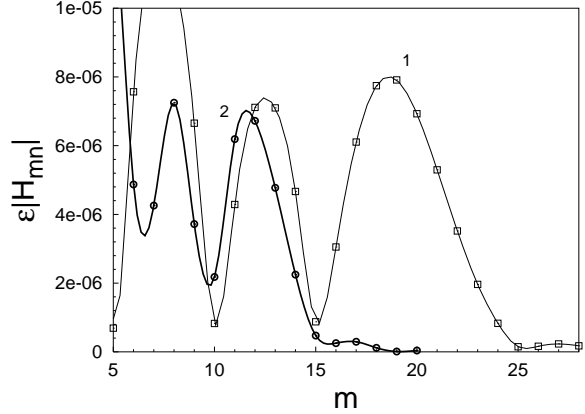} 
  \caption{Typical poloidal spectrum of perturbation field $\epsilon |H_{mn}|$
    at the magnetic surfaces $q=3$ in the TEXTOR-DED: curve 1 describes
    $b_m(r,\pi)$, curve 2 corresponds to numerically calculated
    $H_{mn}(\psi)$. Parameters of the plasma are $B_t=1.9$~T, $I_p=$~382~kA,
  $\beta_{pol}=0.5$, $a=44.7$ cm, $R_0=175$ cm, the DED current, $I_d$=15 kA.} 
  \label{spectra_DED}
\end{figure}
It is shown as curve 1 in Fig.~\ref{spectra_DED} for the toroidal mode
$n=4$. The radial decay of the modes $|b_{mn}|$ has a power law
$|b_{mn}(r,\theta)| \propto (r/r_c)^m$, and correspondingly the full radial 
perturbation magnetic field, $B_r \propto (r/r_c)^{\gamma_n}$, with the
exponent, $\gamma_n \approx m_0-1= 19$, which is twice larger than the 
corresponding exponent $\gamma_n$ for the ED of Tore Supra (see
Eq. (\ref{B_r_gamma})). It means that the perturbation field in the TEXTOR-DED
does not penetrate into the plasma much deeper than in the case of the ED of
Tore Supra. 

On the other hand the transformation from geometrical to intrinsic
coordinates for the TEXTOR-DED case, modifies the perturbation field spectrum,
$\epsilon |H_{mn}|$, in a completely different way. An example of this spectrum 
is shown in Fig.~\ref{spectra_DED} as curve 2.
The central mode number $m_0$ of the poloidal spectrum in the geometrical
space is about $m_0 \approx 20$ and the width $\Delta m \approx 10$. In
intrinsic coordinate the central mode is shifted to $m_0^* \approx 12$, and
the width $\Delta m^* \approx 6 $.
 
As was shown in
Refs. \cite{Abdullaev_etal_99,Finken_etal_99,Finken_etal:2005} this spectrum has the
following asymptotics  
\begin{align}   \label{H_mn_asym2}
H_{mn}(\psi) \approx &(-1)^{m}\beta_1^{-1} B b_{m'(m) n}\left(r,\pi=\pi\right),
  \cr 
  &m'(m)= \frac{m + x_c \left(\beta_3|m/2\right)^{1/3}}{\beta_1},
\end{align}
where $\beta_1$ and $\beta_3$ are the first and the third derivatives of
$\theta$ with respect to $\vartheta$ taken on the high field side
$\vartheta=\pi$ \cite{Abdullaev:2006,Finken_etal:2005}. We note that $\beta_1
<1$. In this case the central mode $m_c^*$ is shifted to
the lower mode number $m_c^* = m_c \beta_1$, and the width of the spectra is
decreased, $\Delta m^* \approx \Delta m \beta_1$. 

Thus the analytical analysis made in Secs. ~\ref{Determ_Spectra} and ~\ref{Structure_compareTEXTOR} 
show that the effect of the perturbation significantly changes when the coils are moved from one side 
of the torus to the opposite one. If the TEXTOR-DED coils were located on the low-field-side with the same 
coil design the effect of the perturbation would be drastically increased: the group of modes of $b_m$ 
centered near $m_c \approx 20$ would be transfered to the mode group with the central mode number $m_c^* \approx \gamma m_c  \approx 36$ and the width $\Delta m^* \approx \gamma \Delta m \approx 18$ which however are not resonant at the plasma edge.
The resonant modes of $H_{mn}$ with $m \approx 10- 14$ would be formed from the group of modes of $b_m'$ with $m'=5 - 8$.
The radial decay of these modes is much weaker, i.e., $H_{mn} \propto r^\gamma, (\gamma = 5-8)$, and they penetrate much 
deeper into the plasma. In the case of Tore Supra the move of the ED coils to the opposite side would significantly weaken 
the effect of perturbation. The resonant modes $H_{mn}$ with $m=14-20$ will be formed from the group of modes of $b_m'$ 
with $m' \approx 25-33$, which has a much stronger radial decay $H_{mn} \propto r^\gamma, (\gamma = 25-33)$.

In reality the coils for the Tore Supra ED as well as the TEXTOR-DED are designed to have a group of modes near the central mode perturbation, $m_c^*$, which are resonant to field lines near the $q=3$ surface of the plasma edge.

In addition, from Eqs. (\ref{H_mn_asym}) and (\ref{H_mn_asym2}) the following roughly 
determined radial decay laws for the perturbation modes can be derived: $H_{mn}(\psi) \propto r^{m/\gamma_1}, (\gamma_1 >1)$ 
for the ED of Tore Supra and $H_{mn}(\psi) \propto r^{m/ \beta_1}, (\beta_1 <1)$ for the TEXTOR-DED, i.e., in the first
case at a given poloidal mode $m$, the radially inward decay of the perturbation is much weaker than in the second case. This is most probably related to the location of perturbation coils on the low-field-side and the high-field-side,
respectively, although differences in design between the perturbation coils in Tore Supra ED and TEXTOR-DED can not completely 
be ruled out as a possible explanation.

\section{Structure and statistical properties of the ergodic zone}
\label{Structure}
\subsection{Qualitative estimates of the formation of the ergodic zone at the plasma edge} 

Similarly to the analysis which has been made for TEXTOR-DED \cite{Finken_etal:2005}, a qualitative picture of the formation of the ergodic zone in Tore Supra ED can be obtained by considering the Chirikov parameter \cite{Chirikov_79,Lichterberg_92}, which is determined by the perturbation spectrum and the q($\rho$)-profile.

\begin{figure}[htb]
  \centering 
  \includegraphics[width=7cm]{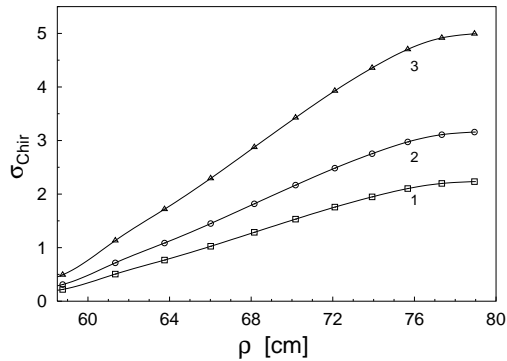} 
  \caption{Chirikov parameter $\sigma_{Chir}$ vs the radial coordinate $\rho$
    for the different perturbation currents: curve 1 corresponds $I_d=4.5$ kA,
    curve 2 $-$ $I_d=9$ kA, and  curve 3 $-$ $I_d=22.5$ kA. The plasma
    parameters are the same as in Fig.~\ref{asymp_spectra}.} 
  \label{sigma_TS}
\end{figure}

This Chirikov parameter, $\sigma_{Chir}$, as a function of the mean radius of
the neighboring resonant magnetic surfaces, $\rho =(\rho_{m+1,n} +
\rho_{mn})/2$, is shown in Fig.~\ref{sigma_TS} for the three different
perturbation currents: $I_d=4.5$ kA (curve 1), $I_d=9$ (curve 2), and 
$I_d=22.5$ kA (curve 3). 
As seen from Fig.~\ref{sigma_TS} the Chirikov parameter, $\sigma_{Chir}$,
grows linearly with the radius, $\rho$, and exceeds the unity (which is the 
criterium for the overlapping of magnetic islands) at $\rho >
\rho_1 \approx 66$ cm for the perturbation current $I_d=4.5$ kA, $\rho >
\rho_2 \approx 64$ cm for $I_d=9$ kA, and  $\rho > \rho_3 \approx 61$ cm for
$I_d=22.5$ kA. Therefore, the field lines are chaotic in the regions, $\rho >
\rho_i$, ($i=1,2,3$), at the corresponding perturbation currents, $I_d=4.5$
kA, $I_d=9$ kA, and $I_d=22.5$ kA, respectively.  

\subsection{Structure of the ergodic zone}

A more precise picture of the onset of chaotic field lines and the formation
of the ergodic zone can be obtained by Poincar\'e sections and laminar plots
of field lines. Below we construct these plots employing the earlier mentioned 
mapping method of integration of field line equations. Furthermore, the
integration step, $\Delta\varphi$, of the mapping along the toroidal angle
$\varphi$, is chosen equal to $2\pi/ns$, with $n=6$ and $s=4$. 

\begin{figure}[ht]
  \includegraphics[width=7cm]{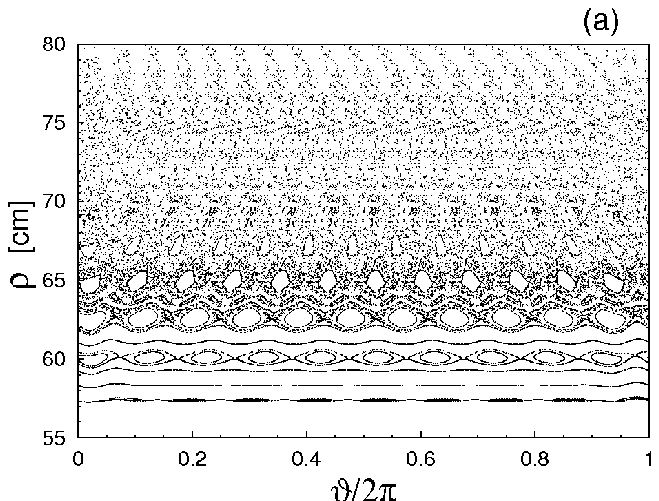}
  \includegraphics[width=7cm]{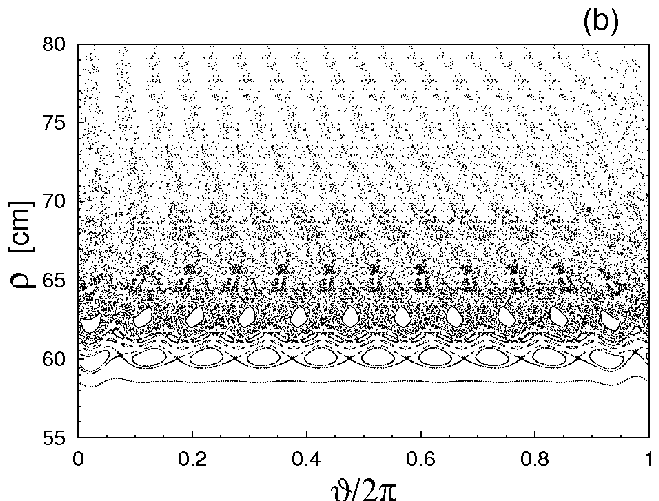}
  \includegraphics[width=7cm]{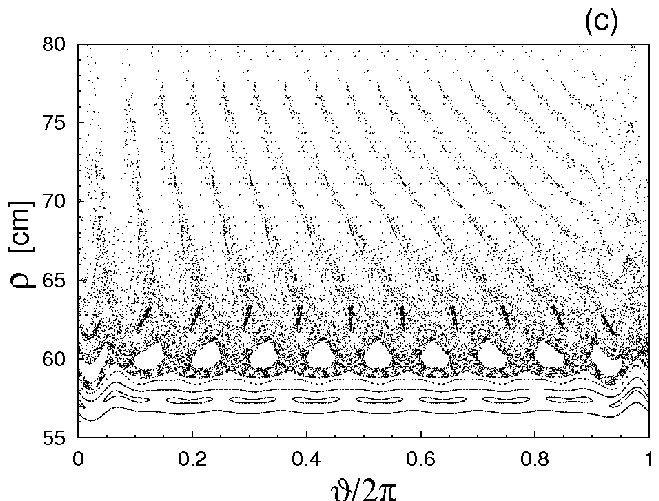}  
  \caption{Poincar\'e sections of field lines in the ($\vartheta, \rho$) plane
    at the three different perturbation currents: (a) $-$ $I_d=4.5$ kA, (b) $-$
  $I_d=9$ kA, and  (c) $-$ $I_d=22.5$ kA. The toroidal modes $n=18$ beside the
    main mode $n=6$ is taken into account into account. The plasma parameters
    are the same as in Fig.~\ref{asymp_spectra}.}   
  \label{Poincare_TS}
\end{figure}

In Figs.~\ref{Poincare_TS}a-c Poincar\'e sections of field lines in the
poloidal plane $\varphi=0$ are plotted for the three different perturbation
currents, $I_d$: (a) $I_d=4.5$ kA, (b) $I_d=9$ kA, and  (c) $I_d=22.5$
kA  for the plasma and ED parameters corresponding to the spectrum of magnetic
perturbations shown in Fig.~\ref{asymp_spectra}. We plotted Poincar\'e
sections in the $(\rho, \vartheta)$ plane which clearly reveal the positions
of the magnetic surfaces. (In this plane the unperturbed field lines are
presented by straight lines, $\rho=$const). 

We can see from Fig.~\ref{Poincare_TS}a that even at the lower level
perturbation current $I_d$=4.5 kA one obtains a rather highly-developed
ergodic zone in the radial region $\rho > \rho_c \approx 63$ cm with open
field lines to the plasma wall. It is formed by the overlap of nine primary
neighboring magnetic islands with the poloidal mode numbers from $m=11$ to
$m=19$ and the toroidal mode $n=6$. There are only two remnants of
KAM-stability islands near the boundary of the ergodic zone with the last
intact magnetic surface $\rho_c \approx 63$ cm. In the radial direction the
ergodization level increases and the remnants of KAM islands drastically
shrink in size. The region of field lines with short wall to wall connection
lengths (white areas in the Poincar\'e plots) grows when one approaches the
plasma edge $\rho=80$ cm. This area is known as a laminar zone.   

With increasing perturbation current $I_d$ the ergodic zone grows radially
inward. At the same time the radial width of the laminar zone increases (see
Fig.~\ref{Poincare_TS}b,c). At the maximal perturbation current
$I_d$=22.5 kA the magnetic island $m=10,n=6$ is overlapped to the ergodic zone
and the radial extension of the laminar zone becomes $\Delta\rho_l \approx 12$
cm. (A more precise way of estimating the laminar zone boundary will be given in
Sect. \ref{Diff} using the radial dependence of the field line diffusion
coefficients).

\begin{figure}[ht]
 \hspace{5cm} (a)  \hspace{6cm} (b) \\
  \includegraphics[width=7cm]{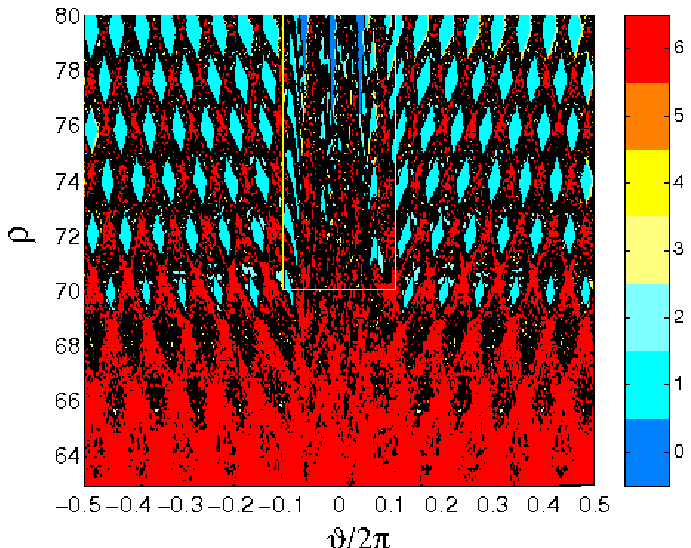} 
  \includegraphics[width=7cm]{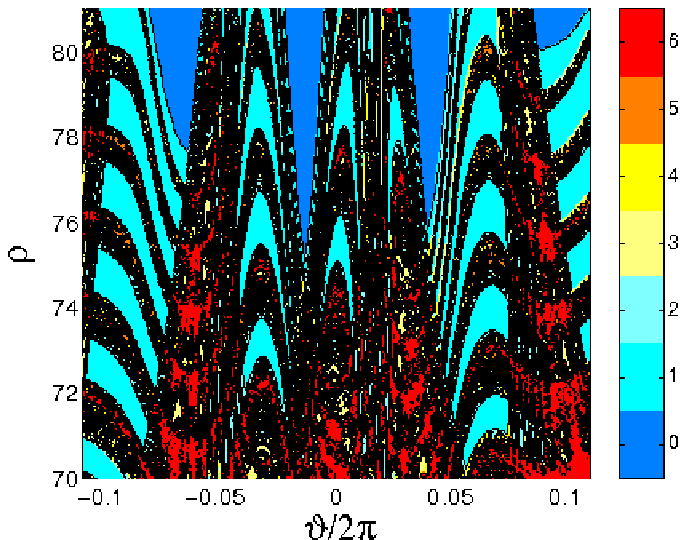}
  \caption{Laminar plots of field lines in the ($\vartheta, \rho$) plane
    at the maximal perturbation current $I_d=22.5$ kA corresponding to
    Fig.~\ref{Poincare_TS}~c: (a) on the whole plane; (b) shows the expanded
    view of the rectangular area in (a). }  
  \label{Laminar_TS}
\end{figure}

The magnetic field lines in the ergodic divertor are a typical example of an
open chaotic scattering system and it is characterized by a certain (fractal)
structure, especially, in the laminar zone (see, e.g.,
\cite{Abdullaev_etal_01,Finken_etal:2005}). The structure of the latter can be
more clearly visualized by the so-called 
laminar plots, i.e., the contour plots of field line wall to wall connection
lengths. Such a plot of connection lengths measured in poloidal turns is
shown in Figs.~\ref{Laminar_TS}a,b in the $(\vartheta, \rho)$ plane at the
maximal perturbation current $I_d$=22.5 kA: (a) shows the plot in the whole
poloidal section and (b) shows the expanded view of the rectangular area at
the low-field-side (LFS) shown in (a). The corresponding Poincar\'e section
was shown in Fig.~\ref{Poincare_TS}c.  

As seen from this figure the laminar zone forms a regular lattice-like
structure in the large poloidal extended area, except the region on the LFS
where field lines are distorted because of the magnetic perturbation which is
localized on this side of the torus. One notices that the light blue ``cells''
in the laminar plot which correspond to the regions of field lines with wall
to wall connection length of one poloidal turn are regularly located along the
poloidal and radial directions. They are positioned between the resonant
radii, $\rho_{mn}$. Their number along the poloidal direction at the given
radius, $\rho$, coincides with the poloidal mode number, $m$, of the resonant
radius, $\rho_{mn}$, located below the ``cells''. The sizes of the ``cells''
grow with increasing radius $\rho$.       

The expanded view of the laminar plot at the LFS shows also the so-called
private flux zone (dark blue areas) of field lines which do not enter into
plasma. Field lines coming from the inner regions of the plasma are connected
to the wall along the elongated stripes (they are seen as black stripes).    

\subsection{Comparison with the TEXTOR-DED}

As seen from the spectra of the magnetic perturbation, $H_{mn}$ shown in
Fig.~\ref{spectra_DED} the ergodic zone at the plasma edge in the TEXTOR-DED
is formed by the interaction of only a few poloidal modes, $11 \leq m \leq 14$,
(12:4 operational mode) in contrast to the ED of Tore Supra ($11 \leq m \leq
20$). On the other hand, in TEXTOR-DED the magnetic perturbation radial decay
has a power-law dependence with the exponent $\gamma$ twice as large as in the
ED of Tore Supra. Because of that one obtains in normal TEXTOR discharges an
ergodized zone of field lines which are weakly connected to the wall. In order
to increase the magnetic flux due to the perturbation field the plasma column
is shifted to the HFS from the center of the torus. \par  
\begin{figure}[ht]
\hspace{12cm}  \begin{large}(b) \end{large} \\
  \includegraphics[width=7cm]{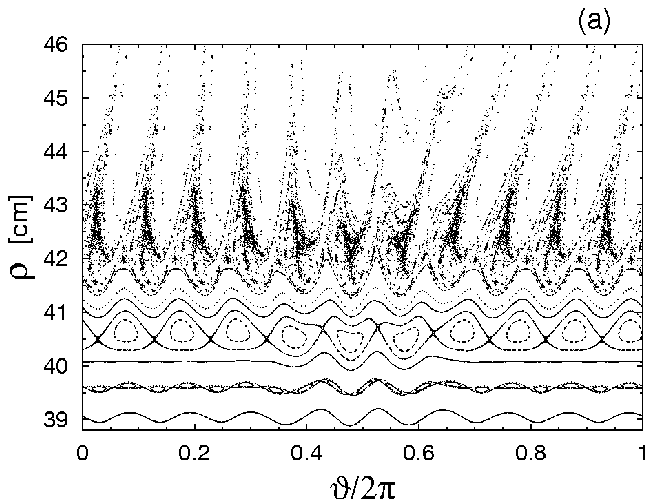}
  \includegraphics[width=6cm]{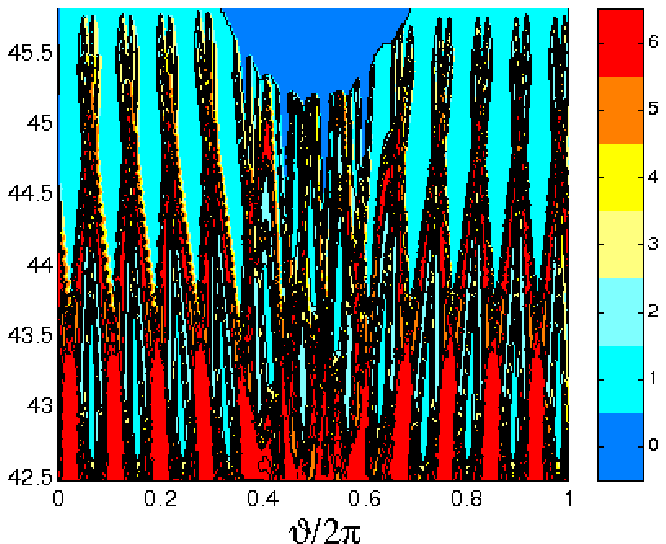}
  \caption{Poincar\'e section (a) and laminar plots (b) of field lines in the
    ($\vartheta, \rho$) plane for the TEXTOR-DED. The plasma parameters are the
    same as in Fig.~\ref{spectra_DED}. The perturbation current $I_d=7$ kA. }  
  \label{Laminar_DED}
\end{figure}

An example of the ergodic zone when the plasma center is shifted over 3 cm from
the center of the torus into the direction of the perturbation coils is shown
in Fig.~\ref{Laminar_DED} where the Poincar\'e section and the laminar plot of 
field lines are displayed. The plasma parameters correspond to the discharge
\#93100: $I_p=$382 kA, $B_t=$ 1.9 T, $\beta_{pol}= 0.5$. \par

As seen from Fig.~\ref{Laminar_DED} in this shot the region of the stochastic
field lines mostly consists of laminar zone with short wall to wall
connection lengths. The width of the laminar zone is radially increased in
expense of the ergodic zone of field lines with long connection lengths.
\par 

\subsection{Magnetic footprints}
\begin{figure}[htbp]
\hspace{3cm} (a) \hspace{7cm} (b) \\
\includegraphics[width=6.7cm]{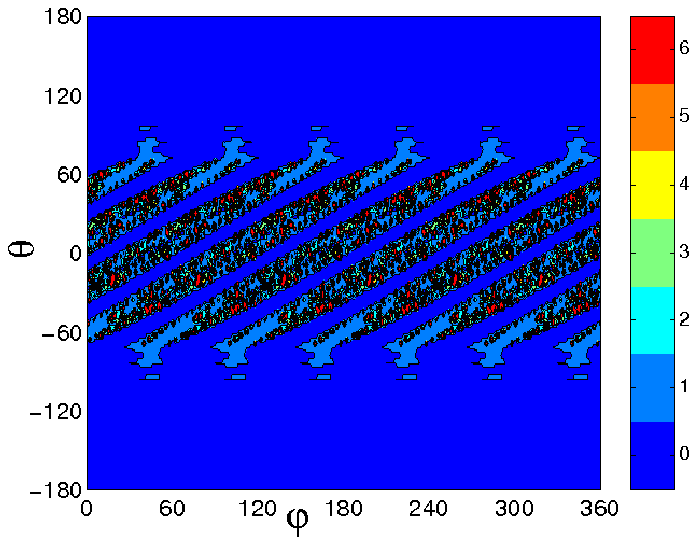}  
\includegraphics[width=6.7cm]{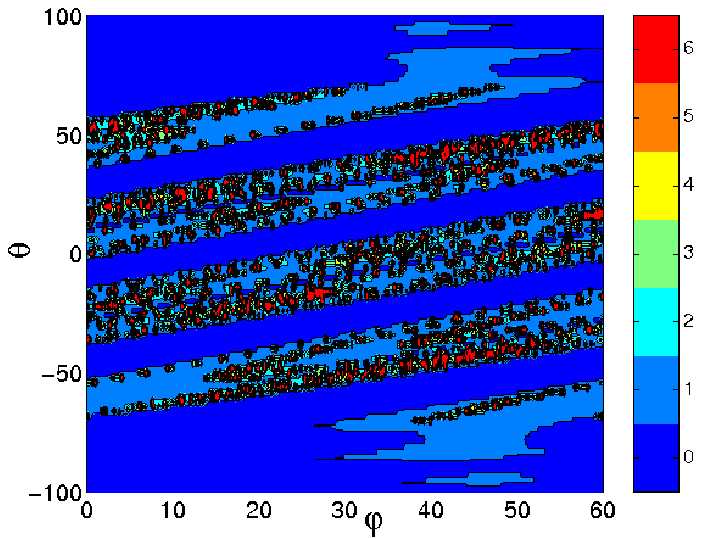} \\
\hspace{5cm} (c) \hspace{7cm} (d) \\
\includegraphics[width=6.7cm]{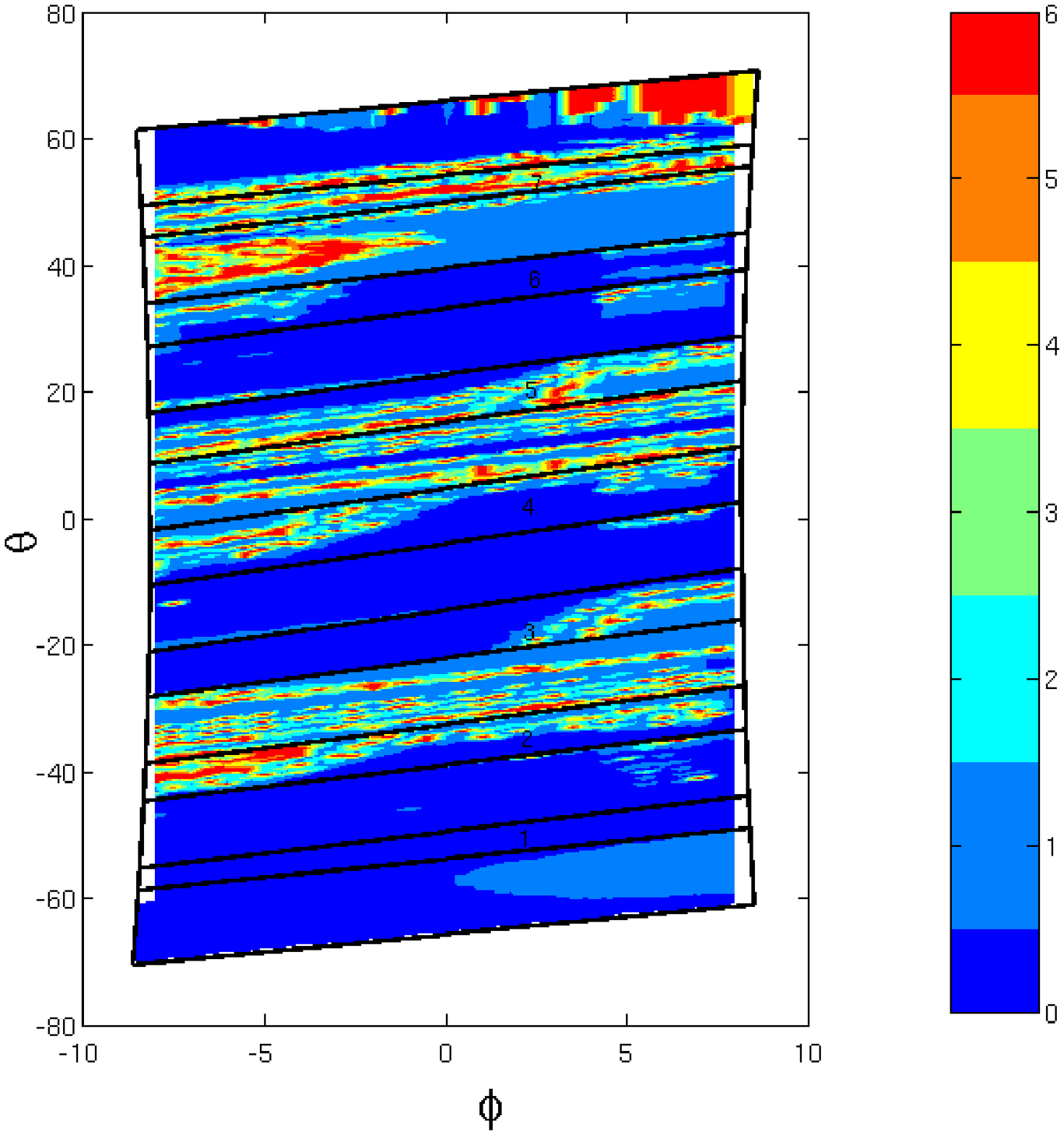} 
\includegraphics[width=6.7cm]{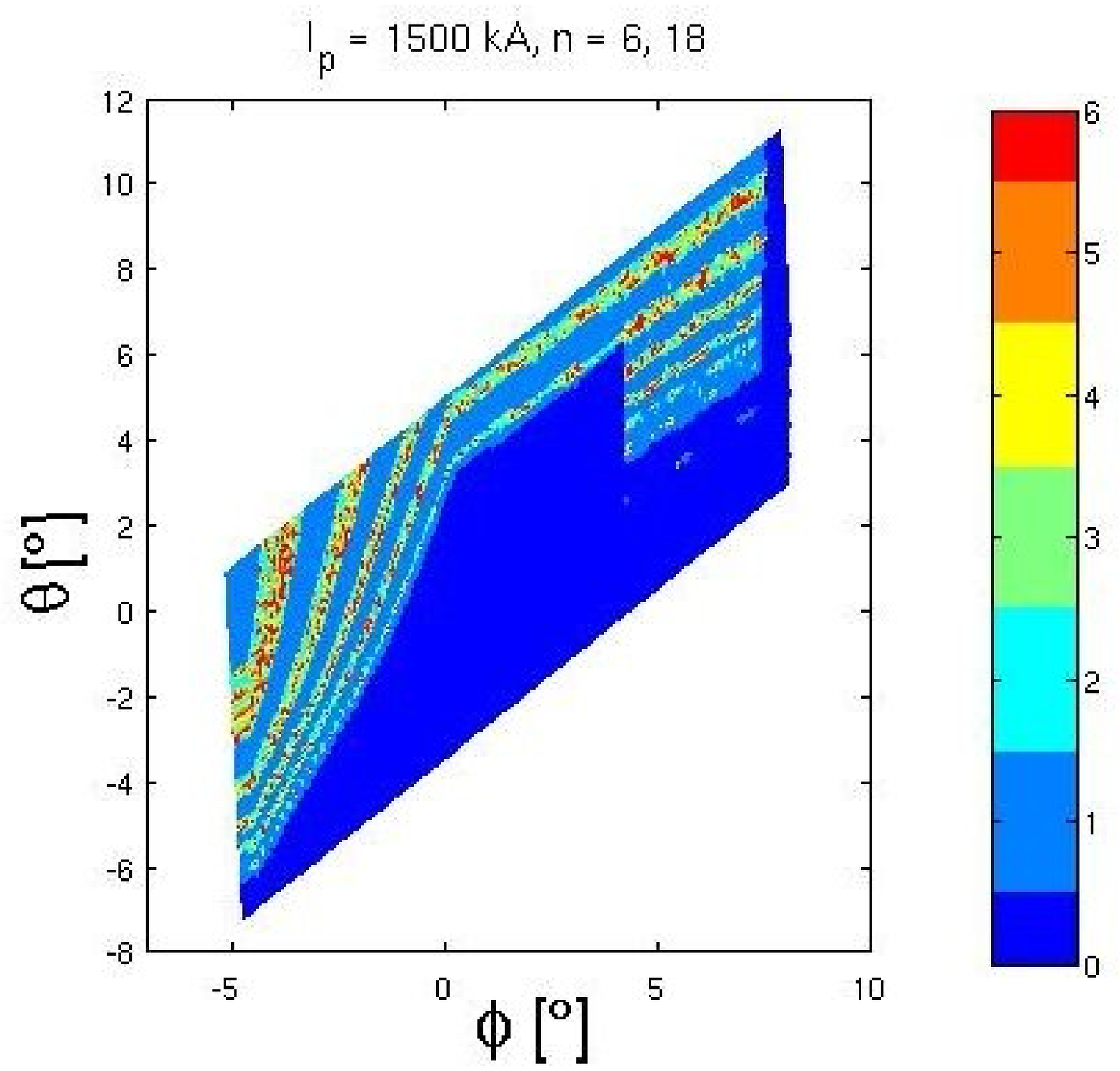} 
\caption{Contour plots of wall to wall connection lengths (magnetic
 footprints) of field lines in the imaginary divertor plate $r_d=80$ cm
 The perturbation current $I_d=22.5$ corresponding to
 Fig.~\ref{Poincare_TS}~c: (a) on the whole plane; (b) shows the expanded
 view of the  region $0<\varphi<60^{\circ}$, $-120^{\circ} < \theta <
 120^{\circ}$. (c) footprint plots on the surface of real divertor plate for ED
  module 1; (d) an expanded view of the above picture on the neutralizer 4.}  
  \label{Footprints_TS}
\end{figure}

Heat and particle deposition patterns on the divertor target plates are
closely linked to the magnetic footprints on the corresponding plates. Since
the geometry of divertor plates of Tore Supra is rather complicated, and in
order to simplify the problem  we study in a first approach the structure of
magnetic footprints on the imaginary divertor plate at the minor radius
$r_d=80$ cm. We present them as contour plots of the wall to wall connection
lengths measured in poloidal turns on the surface of $r_d=80$ cm. The
procedure of corresponding calculations is described in
Refs. \cite{Finken_etal:2005}. \par  

Such a plot of magnetic footprints is given in Fig.~\ref{Footprints_TS} for
the same plasma and ED current parameters as in
Fig.~\ref{Poincare_TS}~c. Fig.~\ref{Footprints_TS}a describes the magnetic
footprints on the whole toroidal surface $r_d=80$ cm, and (b) shows an expanded
view of the region $0<\varphi<60^{\circ}$, $-120^{\circ} < \theta <
120^{\circ}$. As seen from these figures the footprints consists of $n=6$
helical stripes corresponding to the toroidal mode number $n$ as in the case
of the TEXTOR-DED. Field lines coming from the inner plasma region hit the
divertor plate along these helical stripes. Dark blue areas correspond to
the private flux zone from which field lines cannot penetrate into the plasma.
\par 

If, in a more detailed approach, we take the exact geometry of the Tore Supra
divertor plates into account, we need to adjust for two different
aspects. Firstly, we need to modify the criterium used to determine if field
lines hit the wall after each map step. Secondly, the radius $r$ of the field
line mapping starting positions on the ED module surfaces, will be a function
of the toroidal and poloidal angle. A footprint plot in the $(\phi$,$\theta)$ -
plane taking those calculation procedure modifications into account, has been
made for ED module 1 as a whole (shown in Fig.~\ref{Footprints_TS}c) and, with
higher resolution, for neutralizer 4 of the same module
(Fig. ~\ref{Footprints_TS}d). On Fig.~\ref{Footprints_TS}c the edges of the
neutralizer plates have been indicated and the neutralizer plates 
have been numbered. On Fig.~\ref{Footprints_TS}d we notice the pattern of
helical stripes of higher connection length. The bended form of those stripes
is due to the projection on the $\phi$,$\theta$ - plane of the spatial
structure of the neutralizer surface. It is clearly visible that, analogously
to what was found for TEXTOR-DED \cite{Abdullaev_etal_01}, the helical stripes
exhibit a fractal structure, i.e. self-similarity at different spatial scales.

\subsection{Statistical properties of field lines}

In order to estimate the level of radial heat and particles transport
induced by the chaotic field lines one should estimate the statistical
characteristics of those field lines. More precisely, one should study the
radial diffusion coefficients of those field lines and the Kolmogorov lengths
which characterize the degree of divergence of neighboring field lines. Below
we calculate these statistical characteristics and compare them with their
quasilinear estimations. For this we will follow the corresponding procedure
described in Ref.~\cite{Finken_etal:2005}.\par 

\subsubsection{Diffusion of field lines}
\label{Diff}

\begin{figure}[htb]
  \centering 
  \includegraphics[width=7cm]{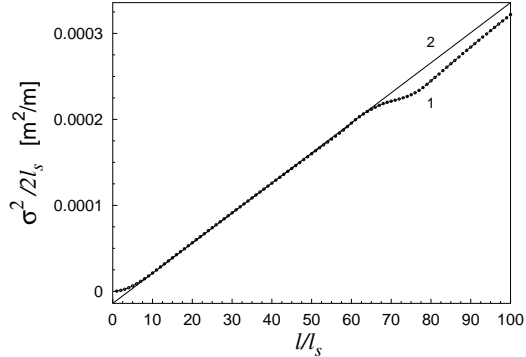} 
  \caption{Determination of a local diffusion coefficient, $D_{FL}$, from the
    linear part of the dependence of the second moment of the radial
    displacement, $\sigma^2(\rho)$, of field lines on their length: curve 1
    corresponds to $\sigma^2(\rho)/2l_s$, vs $l/l_s$; curve 2 describes the
    linear curve   $D_{FL}l/l_s$. Here $l_s$ is a length of a field line per
    one map step. }  
  \label{determination_dfl}
\end{figure}

Consider the second moment of the radial displacement of field lines, 
\begin{equation}   \label{sigma_l}
  \sigma^2(l) = \frac{1}{N} \sum_{i=1}^N \left(\rho_i(l) - \langle \rho (l)
    \rangle \right)^2 ,
\end{equation}
where $l=R_0\varphi$ is a length of field lines. Averaging in (\ref{sigma_l})
is performed over a number $N$ of field lines uniformly distributed at the
initial $\varphi=0$ plane on the given magnetic of radius $\rho$.  
The local radial diffusion coefficient, $D_{FL}$, is defined as \par 
\begin{equation} 
  D_{FL}(\rho) = \sigma^2(l)/2l ,  \nonumber
\end{equation}
at the initial linear growth regime $\sigma^2(l)$, as illustrated in
Fig.~\ref{determination_dfl}.
The diffusion coefficient determined in this way will be a local function of
radius $\rho$.

Diffusion of field lines in the highly developed ergodized zone can also be
estimated in the frame of quasilinear theory. One can show (see,
\cite{Sagdeev_etal:1988,ElskensEscande:2003}) that for the
Hamiltonian system of field lines mentioned in \ref{Determ_Spectra} and extensively described in \cite{Finken_etal:2005} the
distribution function, $f(\psi)$, of field lines along the toroidal flux,
$\psi$ is described by the Fokker-Planck equation 
\begin{equation}   \label{kin_eqn}
  \frac{\partial f(\psi)}{\partial \varphi} = \frac{\partial }{\partial \psi}
  \left( D_M(\psi) \frac{\partial f(\psi) }{\partial \psi} \right), 
\end{equation}
where the diffusion coefficient $D_M(\psi)$ is given by the quasilinear formula
\footnote{Note that the factor 1/2 in the definition of the quasilinear 
diffusion coefficient given in Ref.~\cite{Sagdeev_etal:1988} is absent. This
factor is included into the diffusion equation. }
\begin{equation}   \label{D_M}
  D_M(\psi) = \frac{1}{2} \pi \epsilon^2 \sum_m |mH_{mn}|^2
  \delta\left(\frac{m}{q(\psi)}-n \right). 
\end{equation}
The diffusion coefficient, $D_Q(\rho)$, along the radial coordinate, $\rho$,
is related to $D_M(\psi)$ as 
\begin{equation} \label{D_Q}
  D_Q(\rho) = (R_0^3/\rho^2) D_M(\psi), 
\end{equation}

\begin{figure} [ht]
  \centering 
  \includegraphics[width=7cm]{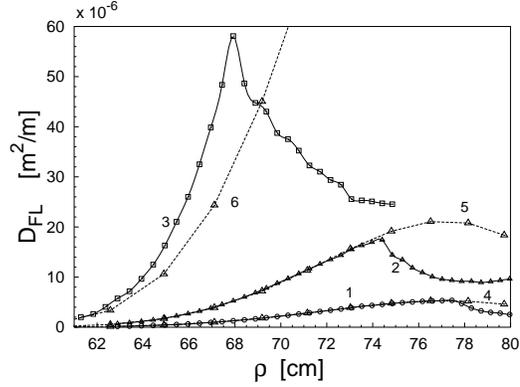} 
  \caption{Radial dependencies of local field line diffusion coefficients,
    $D_{FL}$, of field lines at the three different perturbation currents:
    curves 1 and 4 correspond to $I_d=4.5$ kA, curves 2 and 5 $-$ $I_d=9$ kA,
    and curves 3 and 6 $-$ $I_d=22.5$ kA. Solid curves 1$-$3 describe the
    numerical calculations and dashed curves 4$-$6 describe the quasilinear
    values, $D_Q$.}  
\label{dfl_TS}
\end{figure}

\begin{figure} [ht]
  \centering 
  \includegraphics[width=7cm]{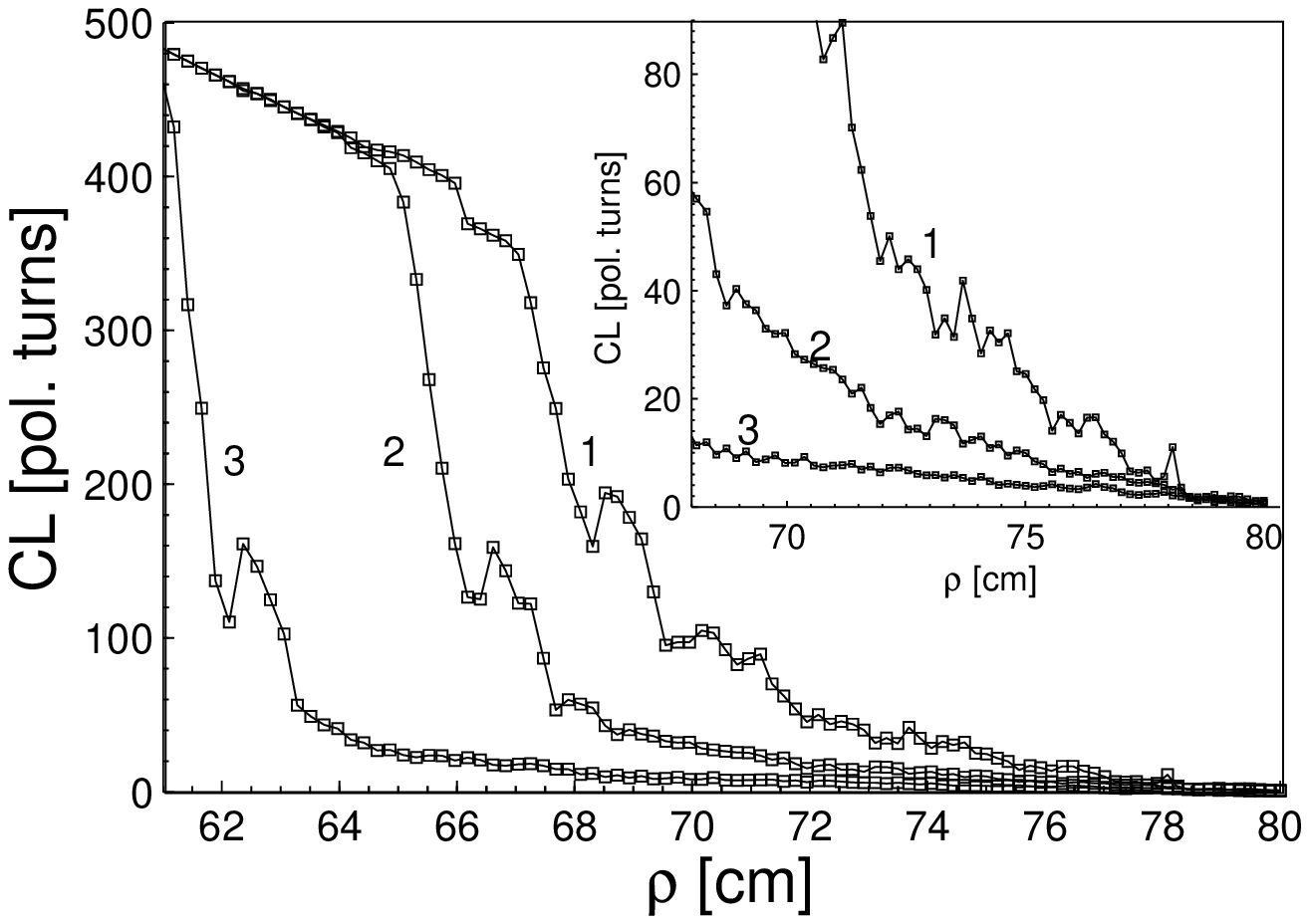} 
  \caption{Radial dependence of the averaged field line connection length,
    measured in poloidal turns, at the three different perturbation currents:
    curve 1 corresponds to $I_d=4.5$ kA, curve 2 $-$ $I_d=9$ kA,
    and curve 3 $-$ $I_d=22.5$ kA. The region nearest to the plasma edge has 
    been enlarged in the upper right corner.}
\label{CL_rad_prof_TS}
\end{figure}

The radial dependencies of the diffusion coefficients 
\footnote{The formulas (\ref{D_M}), (\ref{D_Q}) for the diffusion coefficient
  are given in a singular form. Their actual values should be taken at the
  resonant values of $\rho=\rho_{mn}$ (or $\psi=\psi_{mn}$): $D_M(\psi_{mn}) =
  (1/2) \pi \epsilon^2 q(\psi_mn) m^2 |H_{mn}(\psi_{mn})|^2$. },
$D_{FL}(\rho)$, 
numerically calculated from the field line equations and the corresponding
quasilinear diffusion coefficients, $D_Q(\rho)$ for the three different
perturbation currents, $I_d$, are displayed in Fig.~\ref{dfl_TS}: $I_d=$ 4.5
kA (curves 1 and 4), $I_d=$ 9.0 kA (curves 2 and 5), and $I_d=$ 22.5 kA
(curves 3 and 6), (solid curves 1$-$3 describe $D_{FL}(\rho)$ and dashed curves
4$-$6 $-$ $D_Q(\rho)$). The corresponding radial profiles of the averaged field line
connection lengths are shown in Fig.~\ref{CL_rad_prof_TS}. Again, the averaging was
performed over a number $N$ of field lines uniformly distributed at the
initial $\varphi=0$ plane on the given magnetic of radius $\rho$.

One sees from Fig.~\ref{dfl_TS} that the numerically calculated diffusion
coefficients, $D_{FL}$, steadily grow in certain radial regions, $\rho_{min} <
\rho < \rho_l$. Close examination of the corresponding Poincar\'e sections
given in Figs.~\ref{Poincare_TS} and the averaged connection lenghts shown in
Fig.~\ref{CL_rad_prof_TS} indicates that these radial regions correspond to
the highly developed ergodic zone of field lines possessing the long
connection lenghts with the its lower boundary at $\rho_{min}$ and its
boundary with the laminar zone at $\rho_l$. However, starting from $\rho_l$,
the numerical diffusion coefficients, $D_{FL}$, stop to grow and abruptly go
down, while the quasilinear diffusion coefficients, $D_Q(\rho)$, still
continue to increase. The radius, $\rho_l$, can be called as the lower
boundary of the laminar zone. As seen from Fig.~\ref{CL_rad_prof_TS} in the
laminar zone $\rho > \rho_l$ the field lines have rather short connection
lenghts.  \par 

It is remarkable that for the smaller perturbation currents, $I_d=4.5, 9 $ kA,
the numerical {\em diffusion coefficients, $D_{FL}$, perfectly follow the
quasilinear formula (\ref{D_Q}) in the ergodic zones of field lines with
long connection lengths.} However, they sharply deviate in the laminar zone
with the short connection lengths. This contrasts with the radial profiles of 
diffusion coefficients presented in \cite{Finken_etal:2005} for the case of
the TEXTOR-DED 12:4 operational mode, where the correspondence between
numerical and quasilinear diffusion coefficients in the ergodic zone is
significantly worse. This difference might be explained by noticing that for
the latter case, the applicability of the quasilinear theory is seriously
hampered because the ergodic zone is formed by the overlap of only a few
neighboring magnetic islands. In the ED Tore Supra case, on the contrary, a
higher number of resonances, located closer to each other and thus showing a
larger degree of overlap, result in a much better developed ergodic zone. \par 

On the other hand, for the maximum perturbation current $I_d=22.5$ kA the
situation is different. The difference between the numerical $D_{FL}$ and the
quasilinear diffusion coefficient monotonically increases with the radius
already in the ergodic zone, and the numerical $D_{FL}(\rho)$ exceeds
$D_Q(\rho)$. The reason for such behavior of the $D_{FL}(\rho)$ is not clear
yet and has to be investigated. A first indication might be given by the
Poincar\'e section  of Fig. \ref{Poincare_TS} (c), where the ergodic character
of the radial region  just inward of $\rho_l$ seems to be less pronounced than
in the cases with the smaller perturbation currents. This suggests the
existence of some additional averaged outward drift, due to a deeper reaching
and dominating laminar zone. 

In the case of the TEXTOR-DED the comparison of the numerical $D_{FL}(\rho)$
with the quasilinear diffusion coefficients, $D_Q(\rho)$, is not quite
justified because the ergodic zone in the DED is formed by the interaction of
only three or four magnetic islands. The quasilinear theory of diffusion can
hardly be applied to such an ergodic zone. Particularly, for the case shown in
Fig.~\ref{Laminar_DED} the numerical diffusion coefficient $D_{FL}(\rho)$
reaches the maximal value $3.7 \times 10^{-6}$ m$^2$/m at $\rho=42.95$ cm,
while its quasilinear value is $4.74\times 10^{-5}$ m$^2$/m at the same
radius, i.e., the quasilinear formula overestimates the diffusion coefficient
by a factor 10.

\subsubsection{Kolmogorov lengths} 

The Kolmogorov time is a statistical characteristic of dynamical chaotic
systems which characterizes the time of loss of information on initial
state of the system.  For chaotic field lines it corresponds to the
characteristic decay length of the correlation of neighboring field lines, and
called the Kolmogorov length. The calculation of the latter is based on the
estimations of the Lyapunov exponents. 

The numerical procedure of finding Kolmogorov lengths, $L_K$, for the chaotic
field lines in the ergodic divertor is described in  
\cite{Abdullaev_etal_98,Finken_etal:2005}. It is based on the determination of
the local Lyapunov exponents, $\sigma(\rho)$, describing the degree of
exponential divergency of chaotic field lines launched from a given magnetic
surface of radius $\rho$. The Kolmogorov length, $L_K$, is inverse
proportional to $\sigma(\rho)$ averaged over the given magnetic surface, i.e.,
$L_K = 1/\bar \sigma(\rho)$.

The analytical formula for the Kolmogorov length, $L_K$, has been proposed by
Ghendrih et al. \cite{Ghendrih_etal_92}, which expresses $L_{KQ}$, through the
Chirikov parameter,
\begin{equation}   \label{Kolm_l}
L_{KQ} = \pi q R_0 \left(\pi\sigma_{Chir}/2 \right)^{-4/3}. 
\end{equation}

\begin{figure}[t]
\centering  \hspace{5cm} (a) \\
\includegraphics[width=7cm]{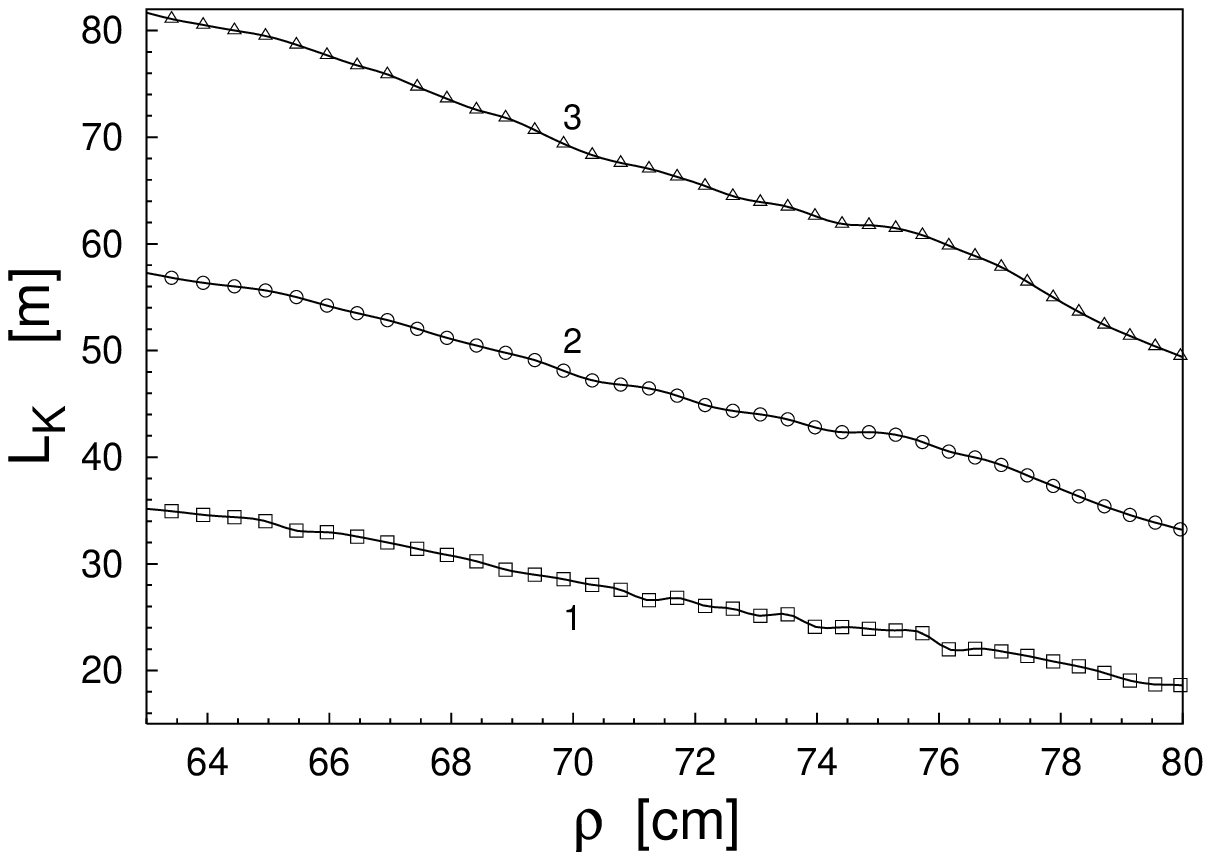} \\\hspace{5cm} (b)\\
\includegraphics[width=7cm]{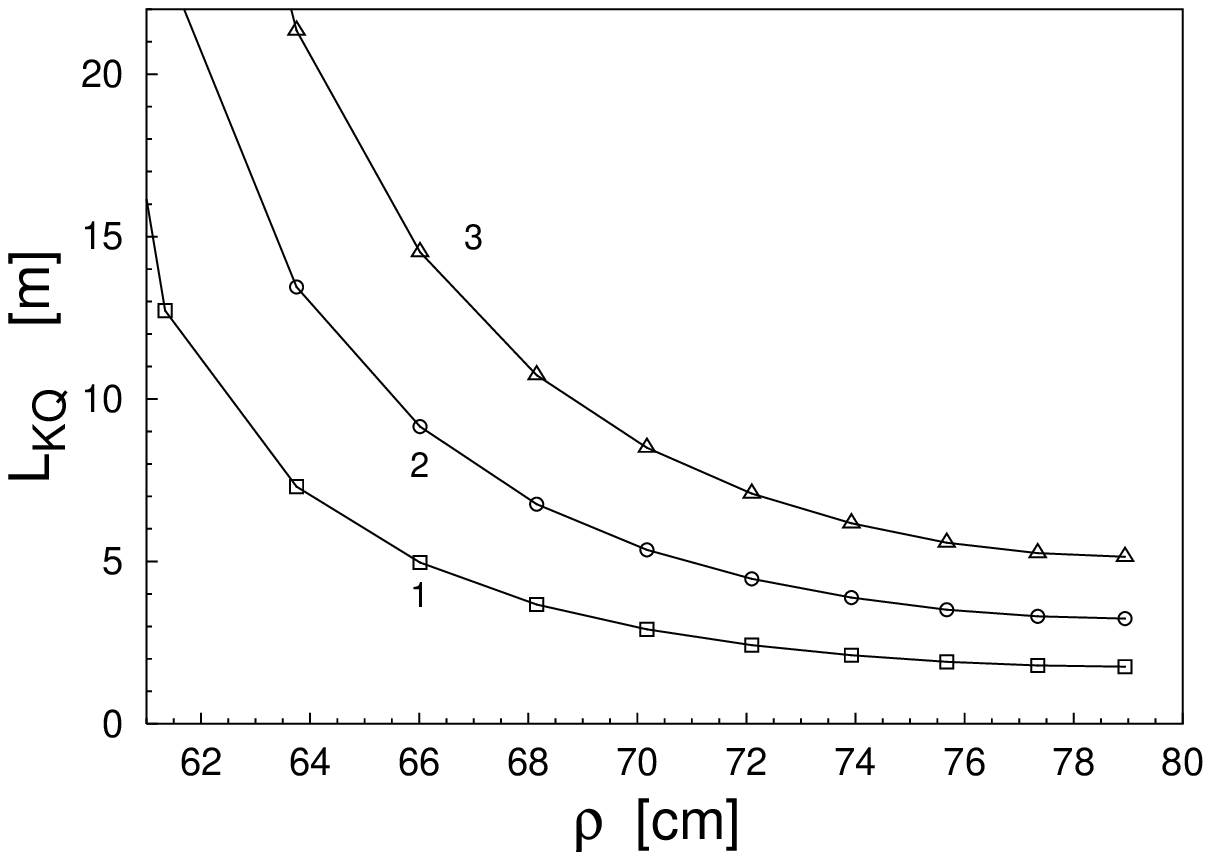} 
\caption{Radial profiles of Kolmogorov lengths, $L_K$, of field lines: (a)
  calculated numerically; (b) according to the quasilinear formula. Curves 1
 to $I_d=4.5$ kA, curves 2 and 5 $-$ $I_d=9$ kA, and curves 3 and 6 $-$
 $I_d=22.5$ kA. }  
\label{L_K_TS}
\end{figure}

The numerically calculated radial profiles of $L_K$ for the chaotic field
lines in the ED of Tore Supra are shown in Fig.~\ref{L_K_TS}a for the three
perturbation currents: curve 1 corresponds to $I_d=$ 4.5 kA, curve 2 $-$ $I_d=$
9 kA and  curve 3 $-$ 22.5 kA. The corresponding profiles of the analytical
formula (\ref{Kolm_l}) are shown in Fig.~\ref{L_K_TS}b. 
The numerically calculated Kolmogorov length, $L_K$, is of the order of the
connection length, $L_c= 2\pi q R_0$, which equal 44.9 m at the $q=3$ magnetic
surface. However, a comparison of Fig.~\ref{L_K_TS}a and b shows that
the numerical Kolmogorov length exceeds the the analytical estimation
(\ref{Kolm_l}) by factors between five or ten. This disperancy is probably 
related to the approximate character of the formula (\ref{Kolm_l}). 
However, the numerical Kolmogorov length is more close to the formula $L_{KQ}
\sim 2\pi q R_0 \sigma_{Chir}^{-4/3}$ given in \cite{Nguyen_etal_97}.

\section{Conclusion}

In summary we have studied the magnetic field structure of the ED of Tore Supra
through the application of the asymptotical and mapping methods developed for
the similar study of the DED of TEXTOR. To this end, an analytical model for
the ergodic divertor of Tore Supra that would  recover the main features of
its magnetic field structure has been proposed. The fine details of the  
magnetic field perturbations, its poloidal and toroidal spectra in the
geometrical coordinate system as well as in straight-field-line coordinates,
the structure of the ergodic and laminar zone, and the field line diffusion
coefficients were investigated. The use of standard code methods similar to
the MASTOC code to study these detailed aspects of the magnetic field in 
the ED would be significantly complicated by the requirement of much higher
computer power or computation times. 

The decay of the magnetic field perturbations for the ED in Tore Supra was
found to exhibit a power--law dependency on the radius, in contrast to an
originally proposed exponential decay law. On the other hand, rigourously
obtained expressions for the width and the central mode of the perturbation
magnetic field poloidal spectrum do coincide with the corresponding
qualitatively obtained original results. An asymptotical formula describing
the transformation law of the poloidal spectra was also obtained. Compared to
TEXTOR-DED, where the ergodic zone at the plasma edge is formed by the
interaction of only a few poloidal modes, the ED has a wide spectrum 
and for a given poloidal mode the radially inward decay of the perturbation
was found to be much weaker. 

For the Tore Supra ED, Poincar\'e sections, laminar plots of chaotic field
lines, as well as the calculation of field line diffusion coefficients showed
a highly-developed ergodic zone at the plasma edge even for relatively low
perturbation currents. Similar to the case of the TEXTOR-DED, the obtained
footprint plots exhibited helical stripes corresponding to the toroidal mode
number and with a fractal structure. It was found that in the ergodic zone
with long wall to wall connection lengths the numerical diffusion coefficients
of field lines are well described by the quasilinear formula, while in the
laminar zone with short wall to wall connection lengths they sharply
deviate. However, for the maximum perturbation current, a different situation
occured. Already in the ergodic zone, the difference between the numerical and
the quasilinear diffusion coefficient, the former exceeding the lather,
monotonically increased with the radius. We have also numerically calculated 
the Kolmogorov lengths of field lines. They have an order of the connection 
lenghts. However, the numerically determined Kolmogorov lenghts exceeds the 
quasilinear prediction by factors between five and ten.

\vspace{1cm}


\begin{thebibliography}{10}

\bibitem{EngelhardtFeneberg_78}
W.~Engelhardt and W.~Feneberg.
\newblock Influence of an ergodic magnetic limiter on the impurity content in a
  tokamak.
\newblock {\em Journal of Nuclear Materials}, 76 \& 77:518--520, 1978.

\bibitem{FenebergWolf_81}
W.~Feneberg and G.~H. Wolf.
\newblock A helical magnetic limiter for boundary-layer control in large
  tokamaks.
\newblock {\em Nuclear Fusion}, 21:669--676, 1981.

\bibitem{Samain_etal_82}
A.~Samain, A.~Grosman, and W.~Feneberg.
\newblock Plasma motion and purification in an ergodic divertor.
\newblock {\em Journal of Nuclear Materials}, 111 \& 112:408--412, 1982.

\bibitem{Gentle_81}
K.~W. Gentle.
\newblock The {Texas} experimental tokamak ({TEXT}) facility.
\newblock {\em Nuclear Technology - Fusion}, 1(4):479--485, 1981.

\bibitem{deGrassie_etal_84}
J.~S. deGrassie, N.~Ohyabu, N.~H. Brooks, and et al.
\newblock Ergodic magnetic limiter experiments on text with a 7/3 resonance.
\newblock {\em Journal of Nuclear Materials}, 128-129:266--270, 1984.

\bibitem{Ohyabu_etal_84a}
N.~Ohyabu, J.~S. DeGrassie, N.~H. Brooks, and et al.
\newblock Preliminary results from the ergodic magnetic limiter experiment on
  the {TEXT} experimental tokamak.
\newblock {\em Journal Nuclear Materials}, 121:363--367, 1984.

\bibitem{Ohyabu_etal_84b}
N.~Ohyabu, J.~S. de{ }Grassie, N.~Brooks, and {et al.}
\newblock Ergodic magnetic layer experiment.
\newblock {\em Nuclear Fusion}, 25(11):1684--1688, 1985.

\bibitem{McCool_etal_89}
S.~C. McCool, A.~J. Wootton, A.~Y. Aydemir, and {et al.}
\newblock Electron thermal confinement studies with applied resonant fields on
  {TEXT}.
\newblock {\em Nuclear Fusion}, 29(4):547--562, 1989.

\bibitem{McCool_etal_90}
S.~C. McCool, A.~J. Wootton, M.~Kotschenreuther,  and {et al.}
\newblock Particle transport studies with the applied resonant fields on text.
\newblock {\em Nuclear Fusion}, 30:167-- 173, 1990.

\bibitem{Deschamps_etal_84}
P.~Deschamps, A.~Grosman, M.~Lipa, and et~al.
\newblock Power exhaust and plasma-surface interaction control in {TORE SUPRA}.
\newblock {\em Journal of Nuclear Materials}, 128-129:38--47, 1984.

\bibitem{Samain_etal_84}
A.~Samain, A.~Grossman, T.~Blenski, G.~Fuchs, and B.~Steffen.
\newblock An ergodic divertor for {TORE-SUPRA}.
\newblock {\em Journal of Nuclear Materials}, 128 {\&} 129:395--399, 1984.

\bibitem{Lipa_etal_88}
M.~Lipa, R.~Aymar, P.~Deschamps, et~al.
\newblock Mechanical design and manufacture of magnetic ergodic divertor for
  the {Tore Supra} tokamak.
\newblock In A.~M. Van{ }Ingen, A.~Nijsen-Vis, and H.~T. Klippel, editors, {\em
  Fusion Technology. Proc. 15th Symp. Utrecht, 1988}, volume~1, pages 874--878,
  Amsterdam and New York, 1989. Elsevier.

\bibitem{EquipeToreSupra_98}
{A. Grosman}.
\newblock Heat flux exhaust in {TORE SUPRA} in ergodic divertor and limiter
  configurations.
\newblock In {\em Fusion Energy 1998. (Proc. 17th Int. Conf. Yokohama)},
  Vienna, 2000. IAEA.
\newblock CD-ROM file EXP4/04 and {\sf
  http://www.iaea.org/programmes/ripc/physics/start.htm}.

\bibitem{Ghendrih_etal_96}
Ph. Ghendrih, A.~Grossman, and H.~Capes.
\newblock Theoretical and experimental investigations of stochastic boundaries
  in tokamaks.
\newblock {\em Plasma Physics and Controlled Fusion}, 38:1653 --1724, 1996.

\bibitem{Grosman_99}
A.~Grosman.
\newblock Review of experimental achievements with stochastic boundaries.
\newblock {\em Plasma Physics and Controlled Fusion}, 41:A185--A194, 1999.

\bibitem{Ghendrih_etal_01}
Ph. Ghendrih, M.~B{\'e}coulet, L.~Costanzo, and et~al.
\newblock Ergodic divertor experiments on the route to steady state operation
  of {Tore Supra}.
\newblock {\em Nuclear Fusion}, 41:1401--1412, 2001.

\bibitem{Ghendrih_etal_02}
Ph. Ghendrih, M.~Becoulet, L.~Colas, , and et~al.
\newblock Progress in ergodic divertor operation on {Tore Supra}.
\newblock {\em Nuclear Fusion}, 42:1221--1250, 2002.

\bibitem{Shoji_etal_92}
T.~Shoji, H.~Tamai, Y.~Miura, and {et al.}
\newblock Effects of ergodization on plasma confinement in {JFT-2M}.
\newblock {\em Journal of Nuclear Materials}, 196 {\&} 198:296--300, 1992.

\bibitem{Evans_etal_89}
T.~Evans, J.~S. Degrassie, H.~R. Garner, { et al.}
\newblock Resonant helical divertor experiments in ohmic and auxiliary heated
  {JIPP T-IIU }plasmas.
\newblock {\em Journal of Nuclear Materials}, 162 {\&} 164:636--642, 1989.

\bibitem{Takamura_etal_87}
S.~Takamura, N.~Ohnishi, H.~Yamada, and T.~Okuda.
\newblock Electric and magnetic structure of an edge plasma in a tokamak with a
  helical magnetic limiter.
\newblock {\em Physics of Fluids}, 30:144--147, 1987.

\bibitem{Takamura_etal_89}
S.~Takamura, Y.~Shen, H.~Yamada, M.~Miyake, T.~Tamakoshi, and T.~Okuda.
\newblock Electric and magnetic structure of tokamak edge plasma with static
  and rotating helical magnetic limiter.
\newblock {\em Journal of Nuclear Materials}, 162-164:643--647, 1989.

\bibitem{Shen_etal_89}
Y.~Shen, M.~Miyake, S.~Takamura, T.~Kuroda, and T.~Okuda.
\newblock Ergodic magnetic limiter experiments in the {HYBTOK-II} tokamak.
\newblock {\em Journal of Nuclear Materials}, 168:295--303, 1989.

\bibitem{Caldas_etal_02}
I.~L Caldas, R.~L. Viana, M.~S. Araujo, and {et al.}
\newblock Control of chaotic magnetic fields in tokamaks.
\newblock {\em Braz. J. Phys.}, 32:980--1004, 2002.

\bibitem{Pires_etal_05}
C.~J.~A. Pires, E.~A.~O. Saettonne, M.~Y. Kucinski, A.~Vannucci, and R.~L.
  Viana.
\newblock Magnetic field structure in the tcabr tokamak due to ergodic limiters
  with a non-uniform current distribution: theoretical and experimental
  results.
\newblock {\em Plasma Physics and Controlled Fusion}, 47:1609--1632, 2005.

\bibitem{Kawamura_82}
T.~Kawamura, Y.~Abe, and T.~Tazima.
\newblock Formation of magnetic islands and ergodic magnetic layers in
  wall-lapping plasma as a non-divertor concept for a reactor - relevant
  tokamak.
\newblock {\em Journal Nuclear Materials}, 111-112:268--273, 1982.

\bibitem{Hattori_etal_84}
K.~Hattori, Y.~Seike, Z.~Yoshida, S.~Hamaguchi, H.~Yamada, N.~Suzuki, K.~Itami,
  Y.~Kamada, N.~Inoue, and T.~Uchida.
\newblock Shrinkage of tokamak current channel by external ergodization.
\newblock {\em Journal of Nuclear Materials}, 121:368--373, 1984.

\bibitem{DED_97}
K.~H. Finken{(ed.)}.
\newblock Special issue: The dynamic ergodic divertor.
\newblock {\em Fusion Engineering and Design}, 37:335--448, 1997.

\bibitem{Finken_etal_05b}
K.~H. Finken, S.~S. Abdullaev, M.~F.~M. {De Bock},  and {et al.}
\newblock Background and initial experiments with the {Dynamic Ergodic Divertor
  on TEXTOR}.
\newblock {\em Fusion Science and Technology}, 47(2):87--96, 2005.

\bibitem{Finken_etal_06b}
K.~H. Finken, S.~S. Abdullaev, M.~F.~M.~{de Bock}, and {et al.}
\newblock Overview of experiments with the {Dynamic Ergodic Divertor on
  TEXTOR}.
\newblock {\em Contributions to Plasma Physics}, 46(7-9):515--526, 2006.

\bibitem{Finken_etal_04}
K.~H. Finken, S.~S. Abdullaev, W.~Biel,  and {et al.}
\newblock The dynamic ergodic divertor in the {TEXTOR} tokamak: plasma response
  to dynamic helical magnetic field perturbations.
\newblock {\em Plasma Physics and Controlled Fusion}, 46:B143--B155, 2004.

\bibitem{Koslowski_etal_06}
H.~R. Koslowski, E.~Westerhof, M.{ de }Bock,  and {et al.}
\newblock Tearing mode physics studies applying the dynamic ergodic divertor on
  {TEXTOR}.
\newblock {\em Plasma Physics and Controlled Fusion}, 48:B53--B62, 2006.

\bibitem{Wingen_etal_06}
A.~Wingen, S.~S. Abdullaev, K.~H., and K.~H. Spatschek.
\newblock Influence of stochastic fields on relativistic electrons.
\newblock {\em Nuclear Fusion}, 46:941--952, 2006.

\bibitem{Finken_etal_07a}
K.~H. Finken, S.~S. Abdullaev, M.~Jakubowski, R.~Jaspers, M.~Lehnen,
  R.~Schlikeiser, K.~H. Spatschek, R.~Wolf, and {the TEXTOR Team}.
\newblock Runaway losses in ergodized plasmas.
\newblock {\em Nuclear Fusion}, 47:91--102, 2006.

\bibitem{Finken_etal_07b}
K.~H. Finken, S.~S. Abdullaev, M.~Jakubowski, and {et al.}
\newblock Improved confinement due to open ergodic field lines imposed by the
  {Dynamic Ergodic Divertor in TEXTOR}.
\newblock {\em Physical Review Letters}, 98:065001, 2007.

\bibitem{Xu_etal_06}
Y.~Xu, R.~R. Weynants, S.~Jachmich, and {et al.}
\newblock Influence of the static {Dynamic Ergodic Divertor} on edge turbulence
  properties in {TEXTOR}.
\newblock {\em Physical Review Letters}, 97:16503, 2006.

\bibitem{Xu_etal_07}
Y.~Xu, M.~{Van Schoor}, R.~R. Weynants, S.~Jachmich, M.~Vergote, M.~W.
Jakubowski, P.~Beyer amd M.~Mitri, B.~Schweer, D.~Reiser, B.~Unterberg, K.~H.
Finken, M.~Lehnen, and {the TEXTOR team}.
\newblock Edge turbulence during the static dynamic ergodic divertor
experiments in {TEXTOR}.
\newblock {\em Nuclear Fusion}, 47(12):1696--1709, 2007.

\bibitem{Evans_etal_04}
T.~E. Evans, R.~A. Moyer, P.~R. Thomas,  and {et al.}
\newblock Suppression of large edge-localized modes in high-confinement
  {DIII-D} plasmas with a stochastic magnetic boundary.
\newblock {\em Physical Review Letters}, 92:235003, 2004.

\bibitem{Evans_etal_05a}
T.~E. Evans, R.~A. Moyer, J.~G. Watkins, and {et al.}
\newblock Suppression of large edge localized modes with edge resonant magnetic
  fields in high confinement {DIII-D} plasmas.
\newblock {\em Nuclear Fusion}, 45:595--607, 2005.

\bibitem{Evans_etal_06}
T.~E. Evans, R.~A. Moyer, K.~H. Burrel,  and {et al.}
\newblock Edge stability and transport control with resonant magnetic
  perturbations in collisionless tokamak plasmas.
\newblock {\em Nature Physics}, 2:419--423, 2006.

\bibitem{Jakubowski_etal_06}
M.~W. Jakubowski, O.~Schmitz, S.~S. Abdullaev, S.~Brezinsek, K.~H. Finken,
A.~Kr{\"a}mer-Flecken, M.~Lehnen, U.~Samm, K.~H. Spatschek, B.~Unterberg,
R.~C. Wolf, and {the TEXTOR team}.
\newblock Change of the magnetic-field topology by an ergodic divertor and the
effect on the plasma structure and transport.
\newblock {\em Physical Review Letters}, 96:035004, 2006.

\bibitem{Jakubowski_etal_07a}
M.~W. Jakubowski, A.~Wingen, S.~S. Abdullaev, K.~H. Finken, M.~Lehnen, K.~H.
Spatschek, R.~C. Wolf, and {the TEXTOR team}.
\newblock Observation of the heteroclinic tangles in the heat flux pattern of
the ergodic divertor at {TEXTOR}.
\newblock {\em Journal of Nuclear Materials}, 363-365:371--376, 2006.

\bibitem{Jakubowski_etal_07b}
M.~W. Jakubowski, M.~Lehnen, K.~H. Finken, O.~Schmitz, S.~S. Abdullaev,
B.~Unterberg, R.~C. Wolf, and {the TEXTOR team}.
\newblock Influence of the dynamic ergodic divertor on the heat deposition
pattern in {TEXTOR} at different collisionalities.
\newblock {\em Plasma Phys. Control. Fusion}, 49:S109--S121, 2007.

\bibitem{Wingen_etal_07}
A.~Wingen, M.~Jakubowski, K.~H. Spatschek, S.~S. Abdullaev, K.~H Finken,
M.~Lehnen, and the {TEXTOR team}.
\newblock Traces of stable and unstable manifolds in heat flux patterns.
\newblock {\em Physics of Plasmas}, 14:042502, 2007.

\bibitem{Nguyen_etal_95}
F.~Nguyen, Ph. Ghendrih, and A.~Samain.
\newblock {\em Calculation of Magnetic Field Topology of Ergodized Zone in Real
  Tokamak Geometry. Application to the Tokamak {TORE-SUPRA} through the
  {MASTOC} Code}.
\newblock Number DFRC/CAD Preprint EUR-CEA-FC-1539. CEA, Cadarache, 1995.

\bibitem{Nguyen_etal_97}
F.~Nguyen, Ph. Ghendrih, and A.~Grosman.
\newblock Interaction of stochastic boundary layer with plasma facing
  components.
\newblock {\em Nuclear Fusion}, 37:743--758, 1997.

\bibitem{Kaleck_etal_97a}
A.~Kaleck, M.~Hassler, and T.~Evans.
\newblock Ergodization of the magnetic field at the plasma edge by the dynamic
  ergodic divertor.
\newblock {\em Fusion Engineering and Design}, 37:353--378, 1997.

\bibitem{Eich_etal_98}
Th. Eich, K.H. Finken, and A.~Kaleck.
\newblock Modelling efforts on the helical near field divertor of the {Dynamic
  Ergodic Divertor (DED) for TEXTOR}.
\newblock {\em Contributions to Plasma Physics}, 38:112--117, 1998.

\bibitem{Abdullaev_etal_99}
S.~S. Abdullaev, K.~H. Finken, and K.~H. Spatschek.
\newblock Asymptotical and mapping methods in study of ergodic divertor
  magnetic field in a toroidal system.
\newblock {\em Physics of Plasmas}, 6:153--174, 1999.

\bibitem{Finken_etal_99}
K.~H. Finken, S.~S. Abdullaev, A.~Kaleck, and G.~H. Wolf.
\newblock Operating space of the {Dynamic Ergodic Divertor for TEXTOR-94}.
\newblock {\em Nuclear Fusion}, 39:637--661, 1999.

\bibitem{Abdullaev_etal_03}
S.~S. Abdullaev, K.~H. Finken, M.W. Jakubowski, et.~al.
\newblock Overview of magnetic structure induced by the {TEXTOR}-{DED} and the
  related transport.
\newblock {\em Nuclear Fusion}, 43:299--313, 2003.

\bibitem{Finken_etal:2005}
K.~H. Finken, S.~S. Abdullaev, M.~Jakubowski, M.~Lehnen, A.~Nicolai, and K.~H.
  Spatschek.
\newblock {\em The structure of magnetic field in the {TEXTOR}-{DED}},
  volume~45 of {\em Energy Technology}.
\newblock Forschungszentrum Julich, Julich, Germany, 2005.
\newblock Avaliable online: {\sf
  http://www.fz-juelich.de/zb/datapool/page/439/00312\verb+_+Finken.pdf}.

\bibitem{Abdullaev_99}
S.~S. Abdullaev.
\newblock A new integration method of {H}amiltonian systems by symplectic maps.
\newblock {\em Journal of Physics A: Math. Gen.}, 32:2745-- 2766, 1999.

\bibitem{Abdullaev_02}
S.~S. Abdullaev.
\newblock The {Hamilton-Jacobi method and Hamiltonian maps}.
\newblock {\em Journal of Physics A: Math. Gen.}, 35:2811-- 2832, 2002.

\bibitem{Abdullaev:2006}
S.~S. Abdullaev.
\newblock {\em Construction of Mappings for Hamiltonian Systems and Their
  Applications}, volume 691 of {\em Lecture Notes in Physics}.
\newblock Springer-Verlag, Berlin Heidelberg, 2006.


\bibitem{Ghendrih_95}
Ph. Ghendrih.
\newblock R{\'e}sonance du divertor ergodique.
\newblock Technical Report Report EUR-CEA-FC-1537, CEA, Cadarache, 1995.

\bibitem{MorozovSolovev_66}
A.~I. Morozov and L.~S. Solov'ev.
\newblock The structure of magnetic fields.
\newblock In M.A. Leontovich, editor, {\em Reviews of Plasma Physics},
  volume~2, pages 1--101, New York, 1966. Consultants Bureau.


\bibitem{Boozer_83}
A.~H. Boozer.
\newblock Evaluation of the structure of ergodic zone.
\newblock {\em Physics of Fluids}, 26:1288--1291, 1983.

\bibitem{Balescu:1988}
R.~Balescu.
\newblock {\em Transport processes in plasmas: 2. Neoclassical transport
  theory}.
\newblock North-Holland, Amsterdam, 1988.

\bibitem{Boozer_92}
A.~H. Boozer.
\newblock Plasma confinement.
\newblock In {\em Encyclopedia of Physical Science and Technology}, volume~13,
  New York, 1992. Academic Press.

\bibitem{Wesson:2004}
J.~Wesson.
\newblock {\em Tokamaks}, volume~48 of {\em Oxford Engineering Science Series}.
\newblock Clarendon Press, Oxford, 3 edition, 2004.

\bibitem{Fedoryuk:1989}
M.~V. Fedoryuk.
\newblock {\em Asymptotic Methods in Analysis}, volume~13 of {\em Encyclopaedia
  of Mathematical Sciences}.
\newblock Springer, Berlin, 1989.

\bibitem{FinkenWolf_97}
K.~H. Finken and G.~H. Wolf.
\newblock Background, motivation, concept and scientific aims for building a
  dynamic ergodic divertor.
\newblock {\em Fusion Engineering and Design}, 37:337, 1997.

\bibitem{Chirikov_79}
B.V. Chirikov
\newblock A universal instability of many-dimensional oscillator.
\newblock {\em Physics Report}, 52:265--379, 1979.

\bibitem{Lichterberg_92}
A.J. Lichterberg and M.A. Lieberman
\newblock {\em Regular and Stochastic Motion 2-nd Ed.}.
\newblock Springer, New York, 1992

\bibitem{Abdullaev_etal_01}
S.~S. Abdullaev, Th. Eich, and K.~H. Finken.
\newblock Fractal structure of the magnetic field in the laminar zone of the
  dynamic ergodic divertor of the torus experiment for technology-oriented
  research ({TEXTOR}-94).
\newblock {\em Physics of Plasmas}, 8:2739--2749, 2001.

\bibitem{Sagdeev_etal:1988}
R.~Z. Sagdeev, D.~A. Usikov, and G.~M. Zaslavsky.
\newblock {\em Nonlinear Physics. From the Pendulum to Turbulence and Chaos},
  volume~5 of {\em Contemporary Concepts in Physics}.
\newblock Harwood Academics, Chur, Switzerland, 1988.

\bibitem{ElskensEscande:2003}
Y.~Elskens and D.~F. Escande.
\newblock {\em Microscopic Dynamics of Plasmas and Chaos}.
\newblock Institute of Physics, Bristol, 2003.

\bibitem{Abdullaev_etal_98}
S.~S. Abdullaev, K.~H. Finken, A.~Kaleck, and K.~H. Spatschek.
\newblock Twist mapping for the dynamics of magnetic field lines in a tokamak
  ergodic divertor.
\newblock {\em Physics of Plasmas}, 5:196--210, 1998.


\end{thebibliography}
\end{document}